\documentclass[sigconf]{acmart}

\usepackage{color}

\usepackage{graphicx}
\usepackage{subcaption}
\usepackage{listings}
\usepackage{pifont}
\usepackage{amsmath}  
\usepackage{enumitem}
\usepackage{fancybox}
\usepackage{multirow}

\usepackage{tabularx}
\newcolumntype{Y}{>{\centering\arraybackslash}X}

\usepackage{appendix}

\usepackage{amsmath,amssymb,amsfonts}                                                       
\usepackage{float} 
\usepackage{textcomp}
\usepackage{xcolor}
\usepackage{setspace}

\usepackage{booktabs}
\usepackage{threeparttable}

\usepackage{url}
\usepackage{framed}

\usepackage{makecell}

\definecolor{codegreen}{rgb}{0,0.6,0}
\definecolor{codegray}{rgb}{0.5,0.5,0.5}
\definecolor{codepurple}{rgb}{0.58,0,0.82}
\definecolor{backcolour}{rgb}{0.95,0.95,0.92}

\lstdefinestyle{mystyle}{
    backgroundcolor=\color{backcolour},   
    commentstyle=\color{codegreen},
    keywordstyle=\color{magenta},
    numberstyle=\tiny\color{codegray},
    stringstyle=\color{codepurple},
    basicstyle=\ttfamily\footnotesize,
    breakatwhitespace=false,         
    breaklines=true,                 
    captionpos=b,                    
    keepspaces=true,                 
    numbers=left,                    
    numbersep=5pt,                  
    showspaces=false,                
    showstringspaces=false,
    showtabs=false,                  
    tabsize=2
}

\usepackage[linesnumbered,ruled]{algorithm2e}

\usepackage{hyperref}
\hypersetup{hidelinks=true}

\newcommand{\wan}[1]{{\color{cyan!70!blue}{[Wan: #1]}}}
\newcommand{\jiang}[1]{{\color{violet}[Jiang: #1]}}
\newcommand{\zhang}[1]{{\color{orange}[Zhang: #1]}}
\newcommand{\chen}[1]{{\color{cyan!70!yellow}[Chen: #1]}}
\newcommand{\su}[1]{{\color{red}[Su: #1]}}
\newcommand{\hysun}[1]{{\color{green!50!black}[Hysun: #1]}}

\renewcommand{\wan}[1]{}
\renewcommand{\jiang}[1]{}
\renewcommand{\zhang}[1]{}
\renewcommand{\chen}[1]{}
\renewcommand{\su}[1]{}
\renewcommand{\hysun}[1]{}

\usepackage{xspace}
\newcommand{\ourmethod}{\textsc{BinPRE}\xspace}
\newcommand{\ourmethoda}{\textsc{BinPRE$^\#$}\xspace}
\newcommand{\ourmethodb}{\textsc{BinPRE$^-$}\xspace}
\newcommand{\ourmethodc}{\textsc{BinPRE$^*$}\xspace}
\newcommand{\ourmethodd}{\textsc{BinPRE$^{**}$}\xspace}
\newcommand{\ourmethode}{\textsc{BinPRE$^{+}$}\xspace}
\newcommand{\code}[1]{\texttt{\small{#1}}}

\newcommand{\boofuzz}{\textsc{Boofuzz}\xspace}
\newcommand{\peach}{\textsc{Peach}\xspace}
\newcommand{\wireshark}{\textsc{Wireshark}\xspace}
\newcommand{\Polyglot}{\textsc{Polyglot}\xspace}
\newcommand{\AutoFormat}{\textsc{AutoFormat}\xspace}
\newcommand{\Tupni}{\textsc{Tupni}\xspace}
\newcommand{\BinaryInferno}{\textsc{BinaryInferno}\xspace}
\newcommand{\DynPRE}{\textsc{DynPRE}\xspace}

\newcommand{\Polyglota}{\textsc{Polyglot$^{+}$}\xspace}
\newcommand{\AutoFormata}{\textsc{AutoFormat$^{+}$}\xspace}
\newcommand{\Tupnia}{\textsc{Tupni$^{+}$}\xspace}

\newcommand{\cf}{\hbox{\emph{cf.}}\xspace}

\newcommand{\eg}{\hbox{\emph{e.g.}}\xspace}
\newcommand{\ie}{\hbox{\emph{i.e.}}\xspace}

\newcommand{\wrt}{\hbox{\emph{w.r.t.}}\xspace}

\AtBeginDocument{
  }

\let\OLDthebibliography\thebibliography
\renewcommand\thebibliography[1]{
	\OLDthebibliography{#1}
	\setlength{\parskip}{3pt}
	\setlength{\itemsep}{0pt plus 0.3ex}
}

\makeatletter
\let\@authorsaddresses\@empty
\makeatother

\setcopyright{none}
\renewcommand\footnotetextcopyrightpermission[1]{} 

\acmConference[Anonymous Submission to ACM CCS '24]{Proceedings of the 31st ACM Conference on Computer and Communications Security}{October 14--18,2024}{Salt Lake City, U.S.A}

\begin{document}

\title{\ourmethod: Enhancing Field Inference in Binary Analysis Based Protocol Reverse Engineering }

\author{Jiayi Jiang}
\affiliation{
  \institution{Shanghai Key Laboratory of Trustworthy Computing, East China Normal University}
  \city{Shanghai}
  \country{China}}
\email{jyjiangsunny@gmail.com}

\author{Xiyuan Zhang}
\affiliation{
  \institution{Shanghai Key Laboratory of Trustworthy Computing, East China Normal University}
  \city{Shanghai}
  \country{China}}
\email{zxy6538@gmail.com}

\author{Chengcheng Wan}
\affiliation{
  \institution{Shanghai Key Laboratory of Trustworthy Computing, East China Normal University}
  \city{Shanghai}
  \country{China}}
\email{ccwan@sei.ecnu.edu.cn}

\author{Haoyi Chen}
\affiliation{
  \institution{Shanghai Key Laboratory of Trustworthy Computing, East China Normal University}
  \city{Shanghai}
  \country{China}}
\email{haoyichense@gmail.com}

\author{Haiying Sun}
\affiliation{
  \institution{Shanghai Key Laboratory of Trustworthy Computing, East China Normal University}
  \city{Shanghai}
  \country{China}}
\email{hysun@sei.ecnu.edu.cn}

\author{Ting Su}
\affiliation{
  \institution{Shanghai Key Laboratory of Trustworthy Computing, East China Normal University}
  \city{Shanghai}
  \country{China}}
\email{tsu@sei.ecnu.edu.cn}

\begin{abstract}

Protocol reverse engineering (PRE) aims to infer the specification of network protocols when the source code is not available. Specifically, field inference is one crucial step in PRE to infer the field formats and semantics.
To perform field inference, binary analysis based PRE techniques are one major approach category.
However, such techniques face two key challenges --- (1) the format inference is fragile when the logics of processing input messages may vary among different protocol implementations, and (2) the semantic inference is limited by inadequate and inaccurate inference rules.

To tackle these challenges, we present \ourmethod, a binary analysis based PRE tool. \ourmethod incorporates (1) an instruction-based semantic similarity analysis strategy for format extraction; (2) a novel library composed of atomic semantic detectors for improving semantic inference adequacy; and (3) a cluster-and-refine paradigm to further improve semantic inference accuracy. 
We have evaluated \ourmethod against five existing PRE tools, including \Polyglot, \AutoFormat, \Tupni, \BinaryInferno and \DynPRE. 
The evaluation results on eight widely-used protocols show that \ourmethod outperforms the prior PRE tools in both format and semantic inference. 
\ourmethod achieves the perfection of 0.73 on format extraction and the F1-score of 0.74 (0.81) on semantic inference of types (functions), respectively.
The field inference results of \ourmethod have helped improve the effectiveness of protocol fuzzing by achieving 5$\sim$29\% higher branch coverage, compared to those of the best prior PRE tool.
\ourmethod has also helped discover one new zero-day vulnerability, which otherwise cannot be found.

\end{abstract}

\renewcommand{\shortauthors}{Jiayi Jiang et al.}

\maketitle

\section{INTRODUCTION}

\label{sec:introduction}

Protocol reverse engineering~(PRE) aims to infer the specifications~(\eg, field formats, semantics, and state machines) of network protocols, assuming only the protocol messages and/or the program binaries implementing the protocols are available~\cite{huang2022protocol,narayan2015survey}.
The inferred protocol specifications can be applied in many scenarios like protocol fuzzing (\eg, \peach~\cite{peach} and \boofuzz~\cite{boofuzz}), 
and network traffic analysis and auditing (\eg, \wireshark~\cite{wireshark}).
Therefore, building effective PRE techniques is important.

Specifically, \emph{field inference} is one important and necessary step of PRE.
It includes two major relevant tasks: (1) \emph{format extraction}, \ie, inferring the field boundaries of the input messages, and (2) \emph{semantic inference}, \ie, inferring the corresponding semantics of the fields identified from (1).
Moreover, field inference is the requisite for inferring protocol state machines~\cite{cui2007discoverer}.
Thus, in this paper, we focus on improving the effectiveness of field inference because of the importance in its own right.

In the literature, there are two major approaches to achieving field inference, \ie, network-traffic based~(NetT-based)~\cite{hahn2016operational, ye2021netplier, kleber2018nemesys, luodynpre, chandler2023binaryinferno} and execution-trace based~(ExeT-based) PRE techniques~\cite{caballero2007polyglot,lin2008automatic,cui2008tupni, caballero2009dispatcher}.
The NetT-based PRE techniques take static network traces as input and perform statistical analysis to mine the format characteristics exhibited in the traces.
However, their inference accuracy relies on the availability of high-quality network traces that contain diverse protocol messages, which are usually difficult to obtain in practice.
The ExeT-based PRE techniques~(also termed as \emph{binary analysis based} PRE techniques throughout this paper), on the other hand, are resilient to the quality of input messages.
They can obtain rich runtime semantics by monitoring the executed instructions of the protocol implementations.
However, despite the rich runtime semantics, these techniques in practice still face  two key challenges in achieving effective field inference.

\vspace{2pt}
The \emph{first} challenge is that the analysis strategies of format extraction in classic PRE techniques~\cite{caballero2007polyglot,lin2008automatic,cui2008tupni} are fragile to the actual protocol implementations.
Since these strategies are usually implemented based on some heuristic patterns of the executed instructions,
they may become fragile when the logics of processing input messages vary among different protocols.
As a result, these analysis strategies would significantly suffer from the
\emph{over-segmentation} errors (\ie, introducing spurious field boundaries) and the \emph{under-segmentation} errors (\ie, missing true field boundaries).
These errors degrade the accuracy of format extraction, which we will illustrate in Section~\ref{sec:format extraction}.
Even worse, these errors would further undermine the subsequent task of semantic inference.
Because semantic inference would fail on the inaccurate field segmentation.

\vspace{2pt}
The \emph{second} challenge is that the semantic inference abilities of the classic PRE techniques~\cite{caballero2007polyglot,lin2008automatic,cui2008tupni} are limited due to the inadequate and inaccurate inference rules.
For example, \emph{types} and \emph{functions} are the two important field semantics, commonly used by network traffic auditing tools like \wireshark~\cite{wireshark} and protocol fuzzing tools like \peach~\cite{peach} and \boofuzz~\cite{boofuzz}.
Take \boofuzz as an example, it by default supports \emph{five} and \emph{six} major categories of types and functions, respectively.
However, the classic PRE techniques cannot infer types and only infer four categories of functions with low accuracy. For example, our investigation reveals that these techniques can only infer the \emph{command} field with the F1-score of 0.14, the \emph{delimiter} field with the F1-score of 0.22, and the \emph{length} field with the F1-score of 0.56.
The inaccurate results of semantic inference may further affect the downstream applications of PRE.

\vspace{2pt}
In this paper, we introduce two \emph{key} ideas to tackle the preceding two challenges, respectively.
To mitigate the first challenge, we introduce an \emph{instruction-based semantic similarity analysis strategy} to enhance the classic format extraction.
Our \emph{key} insight is that \emph{the bytes from the same field should have similar semantics}.
Specifically, we use \emph{the sequences of instruction operators} accessing these bytes as the proxy of the semantics.
In this way, the bytes accessed by similar sequences of instruction operators are merged into the same field.
This strategy is simple yet effective in tackling segmentation errors. Because it is resilient to the logic of processing input messages in the different protocol implementations.

\vspace{2pt}
To mitigate the second challenge, we build a novel library composed of atomic semantic detectors to improve the adequacy and accuracy of semantic inference. 
More importantly, based on the preceding inference results, we introduce a \emph{cluster-and-refine strategy} to further improve the accuracy of semantic inference.
The key idea is that \emph{the messages with similar formats can offer useful contextual information (\eg, the data values of the same byte offsets) to refine the inference results}. 
Specifically, we cluster the messages with similar formats based on the results of our format extraction.

\vspace{2pt}
We implemented a binary analysis based PRE tool named \ourmethod to support the application of our ideas.
We evaluated it against 5 state-of-the-art PRE tools on 8 popular extant protocols. 
The experiment results show that \ourmethod outperforms prior PRE tools in both format extraction and semantic inference.
For format extraction, \ourmethod achieves the perfection score and the F1-score of (0.73, 0.86) for format extraction, while the classic PRE tools
\Polyglot~\cite{caballero2007polyglot}, \AutoFormat~\cite{lin2008automatic}, \Tupni~\cite{cui2008tupni}, \BinaryInferno~\cite{chandler2023binaryinferno}, and \DynPRE~\cite{luodynpre} achieve the perfection scores and F1-scores of (0.60, 0.66), (0.59, 0.64), (0.49, 0.68), 
 (0.13, 0.35), and (0.24, 0.54), respectively.
During format extraction, \ourmethod reduces 63$\sim$70\% segmentation errors \wrt these prior PRE tools on 8 protocols.
For semantic inference, \ourmethod achieves the average precision and recall values of (0.72, 0.77) and (0.77, 0.91) for inferring field types and functions, respectively. 
Moreover, the field inference results of \ourmethod have helped improve the effectiveness of protocol fuzzing by achieving 5$\sim$29\% higher branch coverage, compared to those of the best prior PRE tool, and discover one new zero-day vulnerability.

Finally, it is important to note that \emph{almost all} the prior binary analysis based PRE tools are not publicly available. We took significant efforts to re-implement the field inference strategies of these prior PRE tools (including \Polyglot, \AutoFormat and \Tupni), and carefully validated our implementations by replicating the experiments of these tools. Thus, we believe one of our contributions is to make these implementations readily and publicly available for fair comparison and replication.

\vspace{2pt}
In this paper, we have made the following contributions: 
\begin{itemize}[leftmargin=*]
\item We introduce an instruction-based semantic similarity analysis strategy to enhance the classic field segmentation and effectively mitigate the improper segmentation errors in format extraction~(Section~\ref{sec:separator}).

\item We build a novel library of atomic semantic detectors to improve the adequacy of semantic inference~(Section~\ref{sec:speculator}) and introduce a \emph{cluster-and-refine} paradigm to improve the accuracy of semantic inference (Section~\ref{sec:corrector}).

\item We implement a tool \ourmethod to support the application of our ideas.
\ourmethod outperforms the prior PRE tools in field inference and demonstrates its usefulness in improving such downstream tasks as protocol fuzzing (Section~\ref{sec:evaluation}).

\item We have made \ourmethod and the re-implementations of prior PRE tools publicly available at \textit{\url{https://github.com/BinPRE/BinPRE}} to facilitate replication of our experiments and benefit the community for studying binary analysis based PRE techniques.

\end{itemize}

\section{Background and Challenges}

\label{sec:problem definition}

In this section, we give the necessary background of protocol reverse engineering, especially field inference, and illustrate the challenges of classic PRE techniques for field inference.

\begin{figure}[t]
  \centering
  \includegraphics[width=1.0\linewidth]{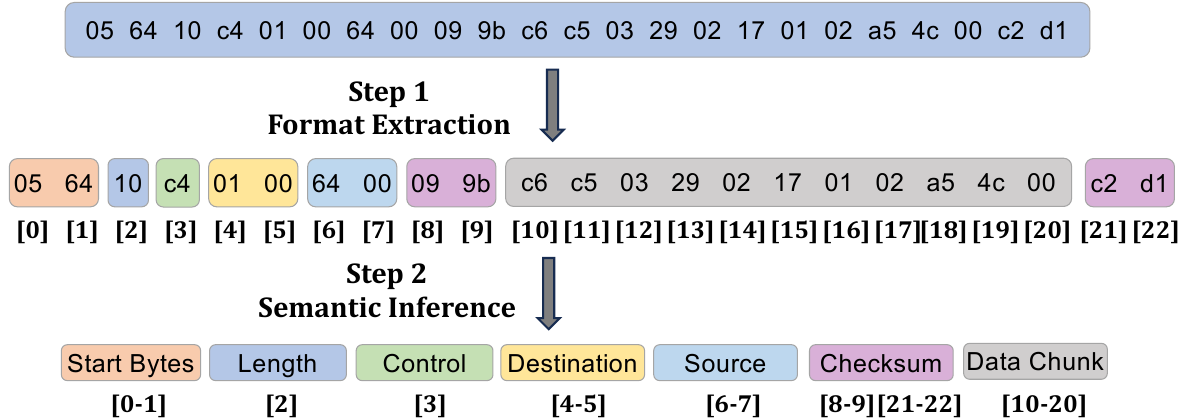}
  \caption{An example message of \texttt{\small{DNP3.0}} protocol~(the fields with different semantics are annotated by different colors).}
  \label{fig:dnp3_example_new}
\end{figure}

\subsection{Background and Definitions}
The main goal of PRE is to infer the protocol fields and the finite state machines of protocols.
This paper focuses on field inference, as it serves as the fundamental pillar of PRE.
In protocol reverse engineering,
\emph{field inference} includes two major tasks: (1) \emph{format extraction}, \ie, inferring the field boundaries, and (2) \emph{semantic inference}, \ie, inferring the corresponding semantics (\eg, \emph{type} and \emph{function}) of the fields.
For example, Figure~\ref{fig:dnp3_example_new} shows a raw binary message in hexadecimal values from  \texttt{\small{DNP3.0}}, a popular communication protocol used in industrial control systems.
To achieve field inference, 
we need to perform format extraction~(Step 1 in Figure~\ref{fig:dnp3_example_new}), and semantic inference~(Step 2 in Figure~\ref{fig:dnp3_example_new}).
Take the second byte (\ie, \texttt{\small{0x10}}) in this message as an example, an ideal field inference should segment this byte into a field, and infer that this field's type is \emph{integer} and its function is the \emph{length} of the message.
In the setting of binary analysis, the tasks of \emph{format extraction} and \emph{semantic inference} can be formulated as follows.

\vspace{2pt}
\noindent\textbf{Format Extraction}.
Let $M$ be a raw message in bytes, \ie, \code{${M}=[b_0, \cdots, b_i, \cdots, b_m]$} ($b_i$ denotes one data byte). Let $INST_M$ be the trace of the executed binary instructions when the program binary processes $M$, \ie, \code{$INST_M=[{inst}_1, \cdots, {inst}_j, \cdots, {inst}_n]$} (${inst}_j$ denotes one executed binary instruction).
By leveraging such classic program analysis as taint analysis~\cite{DTA}, binary analysis based PRE techniques can track which bytes in $M$ have been accessed by which instructions in $INST_M$. 
Based on such information, such techniques can extract field formats. 
For example, one common strategy used by prior PRE techniques~\cite{caballero2007polyglot,lin2008automatic,cui2008tupni} is that, if the instruction ${inst}_j$ accesses the consecutive bytes $[b_k, \cdots, b_l]$ ($0\leq k \leq l \leq m$) in $M$,  the bytes $[b_k, \cdots, b_l]$ is inferred as a field $f$.
As a result, the field boundaries are inferred between the $b_{k-1}$-th and $b_k$-th bytes, and the $b_l$-th and $b_{l+1}$-th bytes in $M$.
This field $f$ is composed of the consecutive bytes from $k$-th to $l$-th in the message $M$.
Here, we can represent $f$ as ${f_{k,l}}=M[k, l]$.
\emph{Format extraction} is such a process of segmenting $M$ into a number of fields.

\vspace{2pt}
\noindent\textbf{Semantic Inference}.
Based on the results of format extraction, \emph{semantic inference} is the process of inferring the semantics of these extracted fields. Specifically, \emph{type} and \emph{function} are the two important field semantic attributes, which are commonly used by protocol fuzzing tools (\eg, \peach~\cite{peach} and \boofuzz~\cite{boofuzz}).
The prior PRE techniques infer the field semantics based on different strategies~\cite{caballero2007polyglot, cui2008tupni, luodynpre, chandler2023binaryinferno}.
For example, in Figure~\ref{fig:dnp3_example_new}, assuming the field ${f_{2,2}}$ is identified, one common strategy used by prior work~\cite{caballero2007polyglot} will infer the function of this field to be \emph{Length} based on some behavioral features of $INST_M$.

\begin{table*}[t]
  \centering
  \small
  \caption{
  F1-scores achieved by the five state-of-the-art PRE tools in performing semantic inference.
  }
  \vspace{-1.0pc}
  \begin{tabular}{c|c|c|c|c|c|c|c|c|c|c|c}
    \toprule
   \multirow{2}{*}{Tool} & \multicolumn{5}{c|}{\emph{\textbf{Type}}} & \multicolumn{6}{c}{\emph{\textbf{Function}}} \\
   \cmidrule(lr){2-6} \cmidrule(lr){7-12}
   &  Static & Integer & Group & Bytes & String & Command & Length & Delim & Checksum & Aligned & Filename\\
   \midrule
    \Polyglot & -  & -  & - & -  & -  & 0.14 & 0.56  & 0.22  & - &  - & -   \\
    \AutoFormat & -  & -  & - & -  &  - & - &  - & -  &  -& -  &  -   \\
    \Tupni &  -  & -   & -  &  -  & -   & -  &  0.56 &  -  & 1.0 & -   &  -   \\
    \BinaryInferno & -  & -  & - &  0 & -  
    & - & 0.50  &  - & - & -  &  -  \\
    \DynPRE &  - &  - & - & -  & -  
    & 0.08 & -  & -  & - & -  &  -  \\
    \hline
  \end{tabular}
  \begin{tablenotes}
      \item * ``-'' denotes that the corresponding semantic (\emph{type} or \emph{function}) is not supported.
      \item * F1-score = 2$\times$precision$\times$recall/(precision+recall)
  \end{tablenotes}
  \label{tab:Qualitative Analysis}
  \vspace{-0.1in}
\end{table*}

\lstset{
  basicstyle=\ttfamily,
  moredelim=[is][\bfseries]{|}{|},
}

\definecolor{mybackground}{HTML}{f8f9fa} 
\definecolor{myblue}{rgb}{0, 0, 1} 
\definecolor{myblue}{HTML}{002fa7} 
\definecolor{mydarkred}{HTML}{43aa8b}
\definecolor{mykeyword}{HTML}{d90429}
\definecolor{highlightcolor}{HTML}{E3170D}

\begin{figure}[t]
\begin{minipage}{0.45\textwidth}
    \begin{lstlisting}[columns=fullflexible, language=C++, caption=Automatak-DNP3 source code snippet.
    , numbers=left, basicstyle=\footnotesize, label=lst:listing1, captionpos=left
    ,frame=single, escapeinside=``, backgroundcolor=\color{mybackground}, keywordstyle = \color{mykeyword}]
bool `\textbf{ShiftableBuffer::\textcolor{myblue}{Sync}}`(){
`\footnotesize{\textcolor{mydarkred}{//2 bytes in $f_{0,1}$ are separately accessed -> Over-seg. errors in the three prior tools}}`
    while (this->NumbytesRead()>1){`\textcolor{mydarkred}{//$f_{0,1}$}`
        if (this->ReadBuffer()[0]==0x05
            && this->ReadBuffer()[1]==0x64){...}}}
            
bool `\textbf{LinkFrame::\textcolor{myblue}{ValidateBodyCRC}}`(const uint8_t* pBody, size_t length){
    length = this->ReadBuffer()[2];`\textcolor{mydarkred}{//$f_{2,2}$}`
    while (length > 0){
        size_t max = LPDU_DATA_BLOCK_SIZE;
        size_t num = (length <= max) ? length : max;
        if (`\textbf{CRC::IsCorrectCRC}`(pBody, num)){`\textcolor{mydarkred}{//$f_{10,20}$, $f_{21,22}$}`
            pBody += (num + 2);`\footnotesize{\textcolor{mydarkred}{//Under-seg. errors in Polyglot}}`
            length -= num;}...}}
            
bool `\textbf{CRC::\textcolor{myblue}{IsCorrectCRC}}`(const uint8_t* input, size_t length){
    ser4cpp::rseq_t buffer(input + length, 2);
    uint16_t crcValue;
    ser4cpp::UInt16::read_from(buffer, crcValue);`\textcolor{mydarkred}{//$f_{21,22}$, $f_{8,9}$}`
    uint16_t CalcCrcValue = 0;
`\footnotesize{\textcolor{mydarkred}{//11 bytes in $f_{10,20}$ are separately accessed -> Over-seg. errors in three prior tools}}`
`\footnotesize{\textcolor{mydarkred}{//8 bytes in $f_{0,7}$ are accessed in a loop -> Under-seg. errors in Tupni}}`
    for (uint32_t i = 0; i < length; ++i){`\textcolor{mydarkred}{//$f_{10,20}, f_{0,7}$}`
        uint8_t index = (CalcCrcValue ^ input[i]) & 0xFF;
        CalcCrcValue = crcTable[index] ^ (CalcCrcValue >> 8);
    }
    CalcCrcValue = ~CalcCrcValue;
    return CalcCrcValue == crcValue;}\end{lstlisting}
\end{minipage}
\vspace{-1.0pc}
\end{figure}

\subsection{Challenges of Format  Extraction}
\label{sec:format extraction}

Format extraction infers the field boundaries of input messages.
However, the strategies of format extraction in classic PRE techniques~\cite{caballero2007polyglot,lin2008automatic,cui2008tupni} are fragile to different protocol implementations.
As a result, these strategies suffer from the \emph{over-segmentation} errors (\ie, introducing spurious field boundaries) and \emph{under-segmentation} errors (\ie, missing true field boundaries).
We use the code snippet (Listing~\ref{lst:listing1}) from \texttt{\small{Automatak-DNP3}}~\cite{dnp3} as example, which is an implementation of the protocol \texttt{\small{DNP3.0}}. 
It processes the input message shown in Figure~\ref{fig:dnp3_example_new}.

\vspace{2pt}
\noindent\textbf{Over-segmentation errors}.
The classic PRE techniques (\eg, \Polyglot~\cite{caballero2007polyglot}, \AutoFormat~\cite{lin2008automatic}, and \Tupni~\cite{cui2008tupni}) assume that different fields have independent data flow from each other
---the bytes belonging to different fields are rarely used in the same instruction.
Thus, they adopt a common strategy for format extraction: \emph{the (consecutive) bytes accessed by one instruction are within the same field} (\cf Section 6 in~\cite{caballero2007polyglot}, Section 3.2.1 in~\cite{lin2008automatic}, Section 3.3 in~\cite{cui2008tupni}).
However, in real-world protocol implementations, the bytes belonging to one field may still be accessed by different instructions, thus leading to over-segmentation errors.
For example, the two bytes of field ${f_{0,1}}$ (\ie, \texttt{\small{0x05}} and \texttt{\small{0x64}}) in the input message, denoting the starting bytes, are accessed by two different instructions (corresponding to lines 4 and 5 in Listing~\ref{lst:listing1}).
As a result, the classic PRE techniques (like \Polyglot, \AutoFormat, and \Tupni) will over-segment field ${f_{0,1}}$ into two fields ${f_{0,0}}$ and ${f_{1,1}}$.
As another example, the eleven bytes of field ${f_{10,20}}$, denoting the data chunk of the input message, are accessed by different instructions (corresponding to line 24 in the loop of lines 23-26 in Listing~\ref{lst:listing1}).
As a result, the classic PRE techniques will over-segment ${f_{10,20}}$ into eleven different fields.

\vspace{2pt}
\noindent\textbf{Under-segmentation errors}.
The classic PRE techniques also implement other heuristic strategies, which may lead to \emph{under-segmentation} errors.
For example, \Polyglot infers the boundaries of variable-length fields based on the \emph{length} fields.
\Polyglot correctly infers field ${f_{2,2}}$ as the \emph{length} field, and identifies its pointer relationship with ${f_{10,20}}$ and ${f_{21,22}}$ (corresponding to lines 12-14).
Therefore, these two fields are erroneously merged into one variable-length field.
As another example, the function \code{IsCorrectCRC} computes the CRC value of the contents in the fields ${f_{0,1}}$, ${f_{2,2}}$, ${f_{3,3}}$, ${f_{4,5}}$, ${f_{6,7}}$ (corresponding to lines 23-26) and checks against the checksum stored in the field ${f_{8,9}}$ (read by line 19).
Since \Tupni merges the consecutive bytes processed by ``mostly'' the same instructions in a loop into one field (\cf Section 3.4$\sim$3.5 in~\cite{cui2008tupni}), the fields ${f_{0,1}}$, ${f_{2,2}}$, ${f_{3,3}}$, ${f_{4,5}}$, ${f_{6,7}}$ are erroneously merged into one field.

Appendix D~(Figure~\ref{fig:dnp3_example_baselines}) gives the format extraction results of the three classic PRE techniques (\Polyglot, \AutoFormat, \Tupni) when processing the input message in Figure~\ref{fig:dnp3_example_new}.

\begin{center}
\doublebox{\parbox{0.46\textwidth}{
    \textbf{Challenge-1}: The classic binary analysis based PRE techniques suffer from the errors of over-segmentation and under-segmentation in performing format extraction.
    This challenge may affect the subsequent task of semantic inference. 
}}
\end{center}

To mitigate the preceding challenge, we introduce an \emph{instruction-based semantic similarity analysis strategy}, which is resilient against the actual protocol implementations (detailed in Section~\ref{sec:separator}).

\begin{figure*}
  \centering
  \includegraphics[width=\linewidth]{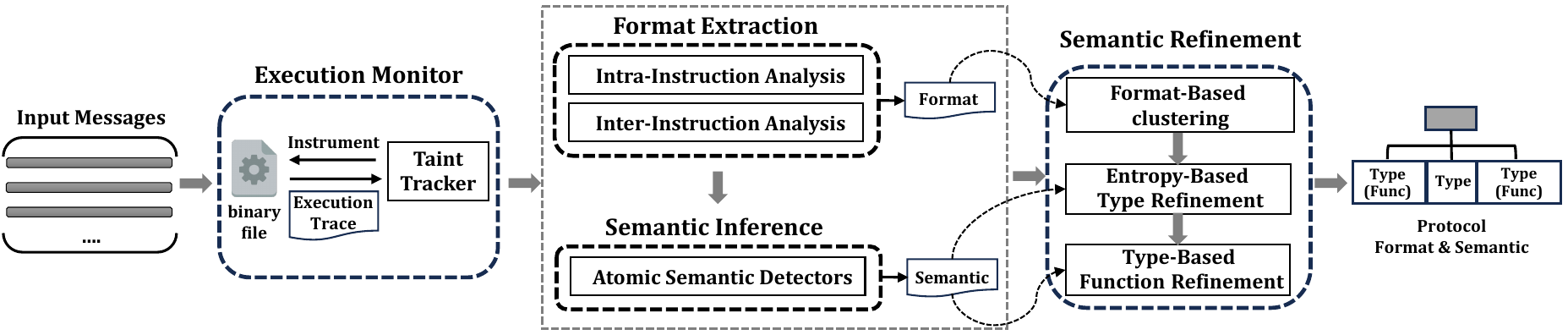}
  \caption{\ourmethod overview}
  \label{fig:worflow}
\end{figure*}

\subsection{Challenges of Semantic Inference}
\label{sec:semantic inference}
Semantic Inference aims to infer the semantics of the fields in a message.
For example, \emph{type} and \emph{function} are the two important field semantics,
which indicate the data type and the meaning of data of a field.
However, the semantic inference abilities of the classic PRE techniques are limited 
due to the inadequate and inaccurate inference rules. 
It affects such the downstream tasks of PRE as network traffic auditing and protocol fuzzing.

To illustrate the limitations, we assess how effective the classic PRE techniques are in supporting
protocol fuzzing, one prominent downstream task of PRE, from the perspective of the adequacy and accuracy of semantic inference.
Specifically, we selected \boofuzz~\cite{boofuzz}, a popular generation-based protocol fuzzing tool, as the reference, and investigated the semantic inference abilities of three classic ExeT-based PRE tools (\Polyglot, \AutoFormat, and \Tupni) and two NetT-based PRE tools (\BinaryInferno and \DynPRE).

\boofuzz by default supports five and six major categories of types and functions~\cite{boofuzzSemantic}.
Table~\ref{tab:Qualitative Analysis} summarizes the semantic inference adequacy and accuracy of the prior PRE tools. 
We can see that none of \Polyglot, \AutoFormat, and \Tupni infers the five types, while \Polyglot only infers three functions and \Tupni infers two functions.
The two NetT-based PRE tools \BinaryInferno and \DynPRE, on the other hand, only infer one function, respectively.
In Appendix A, we give more details of our investigation against \boofuzz as well as another popular protocol fuzzing tool \peach~\cite{peachSemantic}.
Moreover, by replicating the semantic inference of the ExeT-based PRE tools, we find that the ExeT-based PRE tools can only infer the \emph{command} field with an F1-score of 0.14, the \emph{delimiter} field with an F1-score of 0.22, and the \emph{length} field with an F1-score of 0.56 (detailed in Section~\ref{sec:evaluation}).
This indicates that the semantic inference accuracy of the prior PRE tools is far from satisfactory.
It may affect such downstream tasks as protocol fuzzing and network auditing.

\vspace{-0.1in}
\begin{center}
\doublebox{\parbox{0.46\textwidth}{
    \textbf{Challenge-2}:  
    The semantic inference abilities of the classic binary analysis based PRE techniques are limited due to the inadequate and inaccurate inference rules.
    This challenge may affect supporting the downstream tasks of PRE. 
}}
\end{center}

To mitigate the preceding challenge, we build a novel library composed of atomic semantic detectors 
to improve inference adequacy and introduce a \emph{cluster-and-refine} paradigm to improve the accuracy of semantic inference (detailed in Sections~\ref{sec:speculator} and~\ref{sec:corrector}).

\section{\ourmethod DESIGN}

\label{sec:design}

To tackle the challenges in field inference, we introduce \ourmethod, a binary analysis based PRE tool that accurately and comprehensively reverse engineers protocol formats and semantics.

\subsection{Overview}

As shown in Figure~\ref{fig:worflow}, \ourmethod comprises four major modules: \emph{Execution Monitor}, \emph{Format Extraction}, \emph{Semantic Inference}, and \emph{Semantic Refinement}.
Given a protocol message as input, the \emph{Execution Monitor} module tracks its parsing process and records the execution information of the server's instrumented binary. 
The \emph{Format Extraction} module then analyzes the recorded execution information and extracts field format through instruction-based semantic similarity analysis. 
Based on the inferred format and behavioral features within execution information, the \emph{Semantic Inference} module utilizes a library of atomic semantic detectors to identify the semantics of each field. 
Finally, to further improve semantic inference, the \emph{Semantic Refinement} module clusters protocol messages to capture contextual features and refines the semantic inference results.

\subsection{Execution Monitor}
The \emph{Execution Monitor} module incorporates taint analysis~\cite{caballero2007polyglot, lin2008automatic, ma2022automatic} to capture execution information, including data-flow and control-flow when the server parses protocol messages.
It taints protocol message data at the byte level and captures their propagation traces to understand how data is processed by the server program. 

Specifically, the \emph{Execution Monitor} records detailed execution information for each tainted byte, including the taint propagation path, changes in taint value~(\ie, the values of tainted bytes), and the instructions that manipulate the bytes.
It instruments the binary executable of the server at three levels: function, basic block, and instruction. The first two levels capture the control-flow propagation, and the latter tracks the data-flow propagation. 
It pays extra attention to branch conditions and loop iterations, which reflect the dependencies between fields. 
The outputs from all these levels, referred to as behavioral features, are then passed to the \emph{Format Extraction} module and the \emph{Semantic Inference} module.
\vspace{2pt}

\subsection{Format Extraction}
\label{sec:separator}

Given a protocol message and its taint analysis results, \ourmethod uses \emph{an instruction-based semantic similarity analysis strategy} to extract protocol fields and determines field boundaries.
Our key insight is that the bytes from the same field should have similar semantics.
Specifically, the semantics could be approximated by the sequences of instruction operators accessing these bytes.

Algorithm~\ref{alg:alg 1} implements our key insight to achieve format extraction. The algorithm takes as input an input message $M$ and the instruction set $INST_M$ composed of instructions accessing the bytes in $M$.
Recall that $INST_M$ is obtained from the \emph{Execution Monitor} module.
The \emph{Format Extraction} module first conducts \emph{intra-instruction analysis}~(lines 3--8) to identify the adjacent bytes accessed by the same instruction. 
It iterates over all the instructions in $INST_M$ to obtain the field candidates accessed by each instruction.
Specifically, for each instruction, \ourmethod extracts the sequence of bytes from execution information and treats them as a candidate field~(lines 4--5).
It maintains a set of the extracted field candidates $S$ and adds each new identified field candidate to $S$~(line 6).

\begin{algorithm}[t]
\setstretch{1.1}
\footnotesize
\SetAlgoLined
\DontPrintSemicolon
\KwIn{$M$: an input message, $\textit{INST}_{M}$: instructions accessing the bytes in $M$}
\KwOut{$F$: the fields extracted from $M$ }
    \SetKwFunction{FMain}{\textrm{\textbf{\textsc{FormatExtraction}}}}
    \SetKwProg{Fn}{Procedure}{:}{}
    \Fn{\FMain{\text{$M$, $\textit{INST}_{M}$}}}{
        ${S} := \{\}$\hfill $\triangleright$\text{\footnotesize $S$ is the set of field candidates}\\
        \For{$\text{inst}_i$ \text{in} $\textit{INST}_{M}$}{
                ${bs\_inst}_i := \textsc{ExtractByteSeq}\text{(${inst}_i$)}$\\
                ${f\_inst}_i := \textsc{BytesToField}\text{(${bs\_inst}_i$)}$\\
                ${S} := {S}\bigcup \{{f\_inst}_i\} $ \\
        }
        ${L} := \textsc{SortWithOffset}\text{($S$)}$ \hfill $\triangleright$\text{\footnotesize $L$ is a list of field candidates}\\
        ${i} := 0$\\
        \While{$i < (len(L)-1)$}{
            \If{\textrm{ \textbf{\textup{\textsc{SemanticSimilar}}} \textup{(${L}[i]$, ${L}[i+1]$)} }}
            {
                ${L}[i+1] := \textsc{MergeFields}\text{(${L}[i]$, ${L}[i+1]$)}$ \\
                ${L}[i] := Null $ \\
            }
            $i := i + 1$\\
        }
        $F:=RemoveNull(L)$\\
        \textbf{return $F$}\hfill$\triangleright$\text{\footnotesize $F$ is a list of extracted fields}
    }
    \textbf{End Procedure}\\
    
    \SetKwFunction{FSimilar}{\textrm{\textbf{\textsc{SemanticSimilar}}}}
    \SetKwProg{Fn}{Procedure}{:}{}
    \Fn{\FSimilar{\textit{$f_i$, $f_j$}}}{
        $I := \{ inst \in \textit{INST}_{M} \,|\, \textsc{ExtractInstSeq}(inst) \in f_i\}$ \\
        $J := \{ inst \in \textit{INST}_{M} \,|\, \textsc{ExtractInstSeq}(inst) \in f_j\}$ \\
        $m, n := len(I), len(J)$ \\
        $similarity := NW_o(m, n) / max(m, n)$ \\
        \textbf{return $similarity > 0.8$}
    }
    \textbf{End Procedure}
    \caption{Format extraction.}
    \label{alg:alg 1}
\end{algorithm}

To further determine the field boundaries, the \emph{Format Extraction} module then performs \emph{inter-instruction analysis}~(lines 9--19).
It iterates over all the obtained field candidates to examine the semantic similarity between adjacent candidate field candidates~(line 11). 
\ourmethod uses the instruction operators as the proxy of semantics of assembly instructions. Because the instruction operators reflect the core functionalities of these instructions. In practice, \ourmethod extracts the operator sequences accessing each field candidate and adopts the Needleman-Wunsch~(NW) algorithm~\cite{likic2008needleman} (implemented in the procedure \textsc{SemanticSimilar}) to calculate the sequence similarly~(lines 21-25).
Note that Needleman-Wunsch~(NW) algorithm is a classic algorithm for sequence similarity matching scenarios.
When the two operator sequences are similar, their semantics are assumed to be similar.
If the two adjacent field candidates have similar semantics, they will be merged (lines 12--13).

Specifically, given two adjacent field candidates $f_i$ and $f_{j}$, their semantic similarity score $\text{NW}_{o}$ is calculated as
{
\small
\begin{align*}
    {NW}_{o}(i, j) &= \max \left\{
    \begin{array}{ll}
    {NW}_{o}(i-1, j-1) + {C}(I[i], J[j]), \\
    {NW}_{o}(i-1, j) + \code{GAP\_SCORE}, \\
    {NW}_{o}(i, j-1) + \code{GAP\_SCORE}
    \end{array}
    \right.\\
    {C}(a, b) &= \left\{
    \begin{array}{ll}
    \code{MA\_SCORE}, \quad &{if \quad OP(a)=OP(b)} \\
    \code{MISMA\_SCORE}, \quad &{otherwise}
    \end{array}
    \right. 
\end{align*}
}

where $OP(a)$ indicates the operator of instruction $a$; $C(a,b)$ compares whether the operators of the two instructions, \ie, a and b, are the same;\jiang{added more details for C and OP} \code{GAP\_SCORE} (the default value is set as -2) is for penalizing discontinuities between operator sequences; \code{MA\_SCORE} (the default value is set as 1) is for encouraging matches of the same operators; and \code{MISMA\_SCORE} (the default value is set as -1) is for penalizing mismatches of different operators. 
Following the standard convention~\cite{1623879}, the semantic similarity threshold is set as 0.8.
This threshold has been justified by our further evaluation on the effect of different thresholds~(detailed in Appendix C). 
This threshold value ensures a high degree of consistency in deciding the semantic similarity.
By using ${NW}_{o}$, \ourmethod merges semantically similar fields~(lines 12--13).
Finally, it removes all the null fields in $L$ and outputs $F$ as the final format extraction results~(lines 17--18).

\begin{figure}[t]
  \centering
  \includegraphics[width=0.6\linewidth]{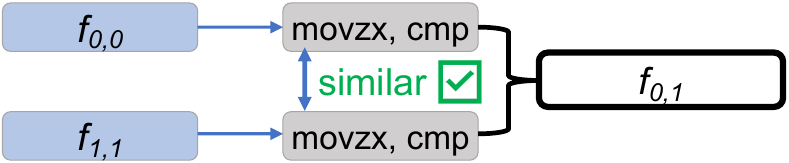}
  \caption{The process of analyzing a field that is prone to the over-segmentation error.}
  \label{fig:illustrative_example1}
\end{figure}

\vspace*{2pt}
To further illustrate the mechanism of \emph{Format Extraction} module, we provide two concrete examples:

\noindent\textbf{Example-1.} Take the first two bytes $[b_0, b_1]$ of the input message in Figure~\ref{fig:dnp3_example_new} as an example. During the intra-instruction analysis, \ourmethod obtains the two field candidates $f_{0,0}$ and $f_{1,1}$ from these two bytes. In Figure~\ref{fig:illustrative_example1}, the gray boxes show the sequences of instruction operators accessing $f_{0,0}$ and $f_{1,1}$, respectively. 
Then, during the inter-instruction analysis, since the sequences of instruction operators accessing $f_{0,0}$ and $f_{1,1}$ are identical (\ie, $[movzx, cmp]$), they have the similar semantic (\ie, the similarity value is 1.0). As a result, \ourmethod merges them into one single field $f_{0,1}$.

\begin{figure}[t]
  \centering
  \includegraphics[width=0.8\linewidth]{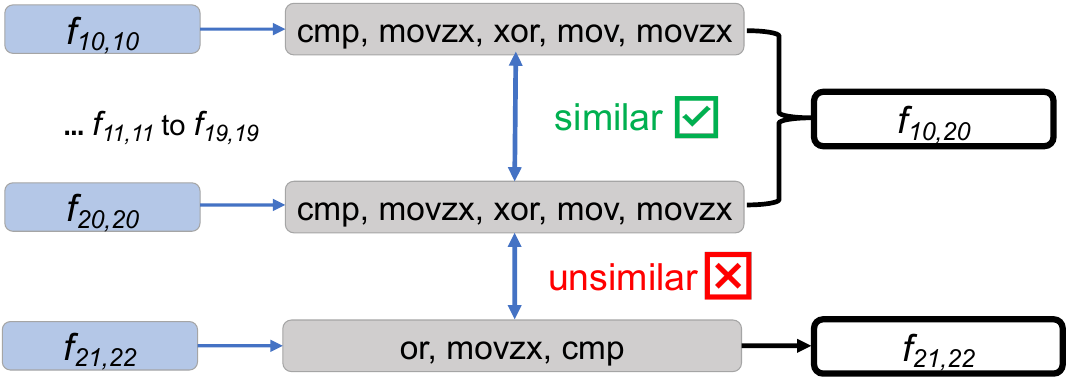}
  \caption{The process of analyzing a field that is prone to the under-segmentation error.}
  \label{fig:illustrative_example2}
  \vspace{-0.5pc}
\end{figure}

\noindent\textbf{Example-2.} Take the bytes $[b_{10}, \cdots, b_{22}]$ of the input message in Figure~\ref{fig:dnp3_example_new} as an example.
As shown in Figure~\ref{fig:illustrative_example2},
all the bytes $[b_{10}, \cdots, b_{20}]$ are processed in the same loop with the identical sequences of instruction operators $[cmp, movzx, xor, mov, movzx]$, while the bytes $[b_{21}, b_{22}]$ are accessed by a different sequence of instruction operators $[or, movzx, cmp]$. As a result, \ourmethod assume the bytes $[b_{10}, \cdots, b_{20}]$ and $[b_{21}, b_{22}]$ have different semantics and thus separates these bytes into two separate fields: $f_{10,20}$ and $f_{21,22}$.

\begin{table*}
\small
  \centering
  \caption{Different semantic types and functions supported by \ourmethod's library of atomic semantic detectors.}
  \vspace{-1.0pc}
    \resizebox{\textwidth}{!}{
    \begin{tabular}{|c|l|l|}
    \hline
    Semantic & \multicolumn{1}{c|}{Attributes} & \multicolumn{1}{c|}{Rules of Atomic Semantic Detectors} \\
    \hline \hline
    Static &  Fixed value across messages & Comparison OPs with a fixed value which yields true \textit{and} without other operations \\
    Integer &  Represent a number of variable length & Arithmetic/bit-wise OPs \textit{or} comparison OPs with multiple consecutive values \\
    Group  &  Comprise a list of available values & Comparison OPs with multiple different constants \\
    Bytes &  Denote a sequence of binary bytes of arbitrary length & Field bytes' shared OPs within a loop\\
    String &  Denote a sequence of text characters & Comparison OPs of consecutive field bytes with the same constant \textit{and} field bytes' shared OPs within a loop\\
    \hline
    Command &  Denote message type & Comparison OPs with a fixed value which yields true \textit{and} corresponding jumping OPs  \\
    Length &  Indicate the length of message or data & Termination OPs of loops, \textit{or} message parsing OPs, \textit{or} pointer-subtraction/counter-decrement OPs\\
    Delim  &  Seperate two fields & Termination OPs of loops \textit{and} delimit OPs of adjacent fields\\
    Checksum &  Verify the integrity of messages or data & Comparison OPs with the output from iterations over multiple consecutive bytes  \\
    Filename &  Identify a file within the file system & Common file naming convention \\
    Aligned &  Align contents to a certain number of bytes & Without functional OPs \\
    \hline
    \end{tabular}
    }
    \label{tab:types_funcs}
\end{table*}

\subsection{Semantic Inference}
\label{sec:speculator}
With behavioral features, the \emph{Semantic Inference} module infers semantic types and functions of each field.
Guided by the execution information from \emph{Execution Monitor}, we construct a novel library of atomic semantic detectors
for inferring each field's \emph{type} and \emph{function}. The semantic detectors utilize two types of information: (1) $I(f)$: the set of instructions accessing field $f$; and (2) $V(f)$: the value of $f$. Note that, all the fields in an input message should have semantic types, and only some fields have semantic functions.

Overall, the \emph{Semantic Inference} module supports five(six) semantic types(functions), which are summarized in Table~\ref{tab:types_funcs}. These semantics align with those supported by \boofuzz~(detailed in Section~\ref{sec:semantic inference}).

\textbf{Static} type has a fixed value and location in the message, regardless of the message content and context.
Its primary purpose is to validate the message and indicate the flags related to protocols, sessions, or messages (\eg, Protocol Version ID). 
\ourmethod regards a field as Static only when 1) the field is compared to a fixed value and the result yields true;
\textit{and} 2) no additional functional operations are performed. Note that, functional operations are instructions other than those whose operators are of the ``mov'' series.

\textbf{Integer} type represents a number recording metadata like data size and length.
Given its numeric nature, \ourmethod regards a field as an Integer when 1) it involves arithmetic or bit-wise operations; \textit{or} 2) it is compared with multiple consecutive values.

\textbf{Group} type comprises a list of static values that encompass all the possible values for the current field. 
It is typically employed to support multiple options, parsing the protocol message based on its value.
Hence, \ourmethod regards a field as Group only when compared with multiple different constants via conditional branches.

\textbf{Bytes} type denotes a sequence of binary bytes of arbitrary length.
It usually serves as data chunks for transmission.
\ourmethod regards a field as Bytes only when all its bytes belonging to the same structure,
identified through identical operations within a loop.

\textbf{String} type is similar to the Bytes type except that it is delimited by a specific delimiter. 
\ourmethod regards a field as String type if the multiple consecutive bytes within this field are compared to the same constant (\ie, a delimiter).
Therefore, \ourmethod distinguishes the String type from the Bytes type by the occurrences of specific consecutive comparison.

\textbf{Command} function indicates the message types, the most critical field in the messages. 
It has different names (\eg, function code, command field, and keyword) in different protocol specifications. 
\ourmethod regards a field as Command only when 1) the field is compared to a fixed value and the result yields true;
\textit{and} 2) when the result yields true, some function jump is immediately triggered.

\textbf{Length} function records the length of the whole or some part of a message. \ourmethod regards a field as Length when 1) it serves as the termination condition of loop structure;
\textit{or} 2) it is retrieved by library APIs (\eg, function \texttt{recv} which takes Length as input); \textit{or} 3) it involves pointer increment and counter decrement operations.

\textbf{Delim} function is inextricably linked to field slicing in protocol implementations, indicating the end of a text protocol field. \ourmethod regards a field as Delim when 1) it serves as the termination condition of a loop;
\textit{and} 2) it delimits its adjacent fields.
Note that, the Delim function is often related to adjacent fields in text protocols, while the Length function is often related to a subsequent block of data or message in binary protocols.

\textbf{Checksum} function verifies the integrity of messages or data, assisting message interaction and communication. 
\ourmethod regards a field as Checksum only when it is compared to the output of a loop which iterates multiple consecutive bytes. 

\textbf{Filename} function stores a file's the path or name. As it is rarely modified or used for control-flow decisions, \ourmethod identifies it by content rather than execution information, considering a field as Filename when it conforms to a common file naming convention.

\textbf{Aligned} function represents the fields that are rarely retrieved by the server, which are mainly for data alignment. When the execution information is missing, such fields are difficult to be identified. \ourmethod regards a field as aligned only when it does not involve any functional operations.

\vspace{-0.5pc}
\begin{figure}[hb]
        \lstset{language=C,
                numbers=left,
                numbersep=5pt,
                tabsize=1,
                xleftmargin=.05\textwidth,
                basicstyle=\footnotesize,
            }
\vspace{-1.0pc}
\begin{minipage}[b]{0.50\columnwidth}
\begin{lstlisting}[columns=fullflexible, language={[x86masm]Assembler}, caption=$I(f)$, numbers=left, basicstyle=\footnotesize, label=lst:listing4,captionpos=left, 
    stringstyle= \color{blue}, frame=single, columns=fullflexible, literate={*}{\hfill}1]
mov qword ptr [rbp-0x40], rsi	
mov rax, qword ptr [rbp-0x40]	
movzx eax, byte ptr [rax]	
movzx edx, byte ptr [rdx]	
shl edx, 0x8	
or eax, edx
'LOOP':
    cmp qword ptr [rbp-0x20], rax	
    movzx eax, byte ptr [rax]	
    xor eax, ecx	
    mov byte ptr [rbp-0x7], al	
    movzx edx, byte ptr [rbp-0x7]	
    ...'repeat for bytes 11-20'
    cmp qword ptr [rbp-0x20], rax   
movzx edx, word ptr [rbp-0x22]	
cmp ax, dx	
\end{lstlisting}
\end{minipage}
\hspace{3pt}
\begin{minipage}[b]{0.45\columnwidth}
\definecolor{mygreen}{RGB}{0,128,0}
\begin{lstlisting}[columns=fullflexible, language={}, caption=$V(f)$., numbers=left, basicstyle=\footnotesize\color{mygreen}, label=lst:listing5,captionpos=left, 
    stringstyle= \color{blue}, frame=single, columns=fullflexible, escapeinside=``]
`\textcolor{codegreen}{$f_{2,2}$}`      V(f):0xb
`\textcolor{codegreen}{$f_{2,2}$}`      V(f):0xb
`\textcolor{codegreen}{$f_{21,21}$}`   V(f):0xc2
`\textcolor{codegreen}{$f_{22,22}$}`   V(f):0xd1
`\textcolor{codegreen}{$f_{22,22}$}`   V(f):0xd1
`\textcolor{codegreen}{$f_{21,22}$}`   V(f):0xc2,0xd100	

`\textcolor{codegreen}{$f_{2,2}$}`      V(f):0xb
`\textcolor{codegreen}{$f_{10,10}$}`   V(f):0xc6
`\textcolor{codegreen}{$f_{10,10}$}`   V(f):0xc6
`\textcolor{codegreen}{$f_{10,10}$}`   V(f):0xc6
`\textcolor{codegreen}{$f_{10,10}$}`   V(f):0xc6

`\textcolor{codegreen}{$f_{2,2}$}`      V(f):0xb
`\textcolor{codegreen}{$f_{21,22}$}`   V(f):0xd1c2
`\textcolor{codegreen}{$f_{21,22}$}`   V(f):0xd1c2
\end{lstlisting}
\end{minipage}

\caption{Assembly instructions for lines 16-28 of Listing~\ref{lst:listing1}}
\label{fig:checksum}
\vspace{-0.7pc}
\end{figure}

\vspace*{2pt}
\noindent\textbf{Example-3.}
Take $f_{21,22}$ in Figure~\ref{fig:dnp3_example_new} as an example. The execution information of $f_{21,22}$ is shown in Figure~\ref{fig:checksum}.
\ourmethod regards its type as Integer, as it involves bit-wise operations (lines 5--6 of Figure~\ref{fig:checksum}).
Meanwhile, the value of $f_{21,22}$ is compared with a value output from a loop that analyzes consecutive bytes~(lines 7--16). Therefore, \ourmethod regards its function as Checksum.

\vspace*{2pt}
\noindent\textbf{Discussion.} To balance accuracy and generality, \ourmethod supports 5 semantic types and 6 semantic functions that are common in protocols and have distinctive features. 
In real-world applications, there might be other field semantics. \ourmethod could easily support them by extending the library of atomic semantic detectors. 
Note that that it is fundamentally impossible for the classic PRE techniques~\cite{caballero2007polyglot,lin2008automatic,cui2008tupni} to achieve similar inference results \wrt \ourmethod, even if we extend  these PRE techniques with our library of atomic semantic detectors~(detailed in Appendix B).

\subsection{Semantic Refinement}
\label{sec:corrector}
The \emph{Semantic Refinement} module enhances the semantic inference results with a \emph{cluster-and-refine} paradigm, incorporating contextual features of fields.
Based on the format extraction results, it first explores the command field position and cluster messages with the most similar formats. 
Within each message cluster, it adopts an entropy-based approach~\cite{shannon1948mathematical}\jiang{revised and added the citation} to characterize field content variation and refine the results of semantic types.
Afterwards, it refines the results of semantic functions, which utilizes the constraints between the semantic functions and the semantic types.

\subsubsection{Format-Based Clustering.} \label{sec:clustering}
With the formats from the \emph{Format Extraction} module, \ourmethod identifies the command field and clusters protocol messages.
This is motivated by the observation that messages with the same command field typically have similar formats~\cite{ye2021netplier}.
Utilizing the similarity of message formats, \ourmethod explores the optimal basis, \ie, command field position, to cluster messages.  
These message clusters contain important contextual features~(\eg, field content variations within each cluster) that would be used to further refine the semantics in \emph{entropy-based type refinement} (\cf Section~\ref{sec:type_refinement}) and \emph{type-based function refinement} (\cf Section~\ref{sec:function_refinement}) steps. 
Note that, this identified command field is also used for refining the earlier inference results.

\begin{figure*}
  \centering
  \includegraphics[width=0.75\linewidth]{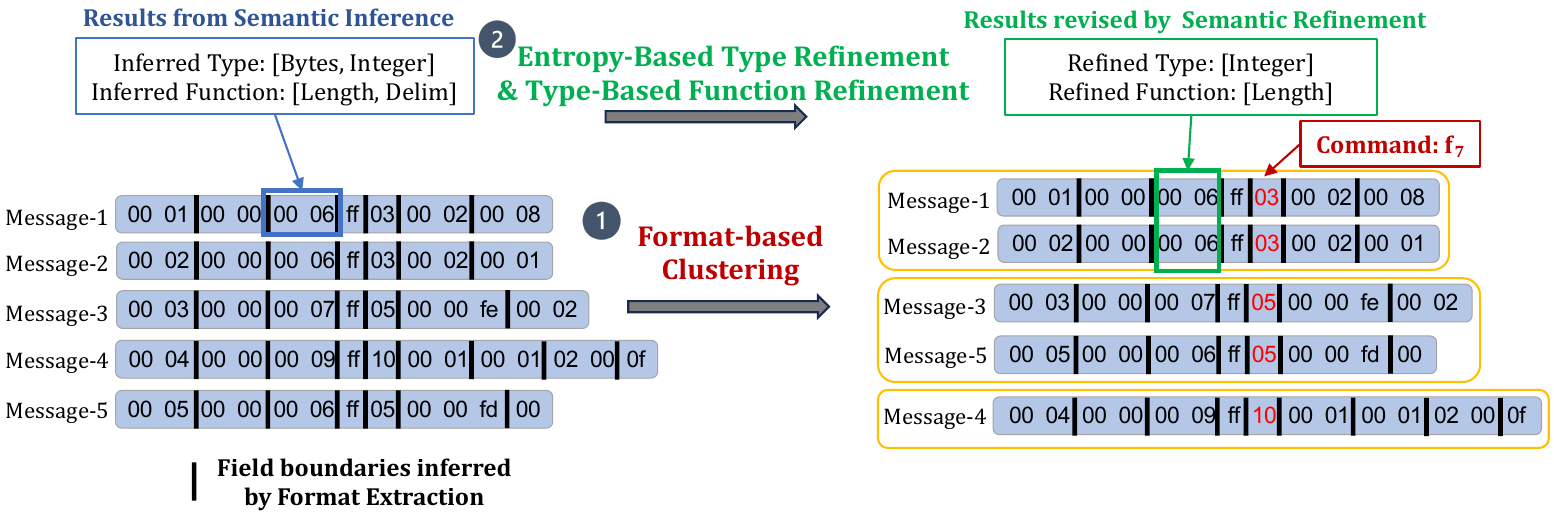}
  \vspace{-0.5pc}
  \caption{An illustrative example of \emph{Semantic Refinement} module.} 
  \label{fig:illustrative_example4}
\end{figure*}

\begin{algorithm}[t]
\setstretch{1.1}
\footnotesize
\SetAlgoLined
\DontPrintSemicolon
\KwIn{\textit{MESSAGES}, \textit{FORMATS}}
\KwOut{\textit{CLUSTERS}: the optimal clustering for messages\\
\quad\quad\quad\,\,\, ${command}_{pos}$: the optimal command position for clustering}
    \SetKwFunction{FMain}{\textrm{\textbf{\textsc{ExploreOptimal}}}}
    \SetKwProg{Fn}{Procedure}{:}{}
    \Fn{\FMain{\textit{MESSAGES}, \textit{FORMATS}}}{
        ${max}_{score} := 0 $ \\
        ${command}_{pos} := -1 $ \\
        \For{${m}_i$ \text{in} \textit{MESSAGES}}{
            ${F}_{i} := \textit{FORMATS}[{m}_i]$ \hfill \text{$\triangleright$ format result of $m_i$} \\
            \For{${b}_j$ \textbf{in} ${F}_{i}$}{
                ${f}_{comm} := \textsc{ExtractField}\text{($b_j$, $b_{j+1}$)}$\\
                ${clusters} := \textsc{Clustering}\text{(${f}_{comm}$, \textit{MESSAGES})}$\\
                ${curr}_{score} := \textsc{AlignScore}\text{($clusters$, \textit{FORMATS})}$\\
                \If{${curr}_{score}$\textgreater  ${max}_{score}$}{
                    ${max}_{score} := {curr}_{score} $\\
                    ${command}_{pos} := {f}_{comm} $
                }
            }
        }
        \textit{CLUSTERS} $:= \textsc{Clustering}\text{(${comm}_{pos}$, \textit{MESSAGES})}$ \\
        \textbf{return} \textit{CLUSTERS}, ${command}_{pos}$
    }
    \textbf{End Procedure}\\
    \caption{Message clustering.}
    \label{alg:alg 2}
\vspace{-0.1pc}
\end{algorithm}

As summarized in Algorithm~\ref{alg:alg 2}, \ourmethod iterates over all message fields and identifies the best basis for format-based clustering. Given a set of messages $\textit{MESSAGES} := \{m_1, m_2, \ldots\}$, and the corresponding set of formats $\textit{FORMATS} := \{F_1, F_2, \ldots\}$ obtained through \emph{Format Extraction} module. For each message $m_i$, \ourmethod iterates over the boundaries in its format results $F_i$~(lines 6--14). 
It extracts a candidate command field $f_{comm}$ from each pair of adjacent boundaries in $F_i$ and clusters messages with it~(lines 7--8). It then calculates the alignment score as the average $NW_f$ score of the format sequences of each message pair within the cluster~(line 9).
Specifically, given two message format sequences, which contain the field boundaries within the messages, \ie, $F_a$ and $F_b$, their format alignment score $\text{NW}_{f}$ is calculated as
{
\small
\begin{align*}
    {NW}_{f}(i, j) &= \max \left\{
    \begin{array}{ll}
    {NW}_{f}(i-1, j-1) + {C}(F_a[i], F_b[j]), \\
    {NW}_{f}(i-1, j) + \code{GAP\_SCORE}, \\
    {NW}_{f}(i, j-1) + \code{GAP\_SCORE}
    \end{array}
    \right.\\
    {C}(m, n) &= \left\{
    \begin{array}{ll}
    \code{MA\_SCORE}, \quad &{if \quad Boundary(m) = Boundary(n)} \\
    \code{MISMA\_SCORE}, \quad &{otherwise}\jiang{revised: ``a,b''->``m,n''}
    \end{array}
    \right. 
\end{align*}
}
where $Boundary(m)$ indicates the offset of boundary m in the message; $ C(F_a[i], F_b[j])$ compares whether the \code{i}th boundary of the message \code{a} and the \code{j}th boundary of the message \code{b}, \ie, $F_a[i]$ and $F_b[j]$, are the same; \code{GAP\_SCORE} (the default value is set as -2) is for penalizing discontinuities between format sequences; \code{MA\_SCORE} (the default value is set as 1) is for encouraging matches of the same boundaries; and \code{MISMA\_SCORE} (the default value is set as -1) is for penalizing mismatches of different boundaries.\jiang{to here.}
In the end of each iteration, the algorithm updates the highest alignment score $max_{score}$ and the optimal clustering basis $command_{pos}$~(lines 10--12).
Finally, \ourmethod accepts the boundary pair of highest alignment score as the Command field, and clusters messages with their Command field value~(lines 16--17).

\subsubsection{Entropy-Based Type Refinement.} 
\label{sec:type_refinement}
Observing that field types mirror the underlying patterns of field value variations, we adopt entropy-based characteristics \jiang{``approaches''->``characteristics''}to refine the results of type inference.

In the \emph{Semantic Inference} module, \ourmethod utilizes execution information to infer semantic types. However, the inference results may not be reliable for the fields that are rarely retrieved. Therefore, \ourmethod utilizes the content variation patterns of a field within the same message clusters to refine these results. It employs an information-theoretic approach to capture the variation patterns of field contents across the messages, with the assumption that messages within the same cluster have similar formats and semantics.
\ourmethod leverages Shannon entropy~\cite{shannon1948mathematical} to characterize the variation patterns. For field $f_{i,j}$, whose boundaries are of offset $i$ and $j$, its Shannon entropy is
{
\small
\begin{align*}
H(f_{i,j}) = -\sum_{v \in C_{i,j}} P(v) \log P(v) \,\,,
\end{align*}
}
where $\text{C}_{i,j}$ is the collection of the $i$-th to $j$-th byte of messages in the Cluster $C$.
The probability of a value $v$ in $\text{C}_{i,j}$, denoted as $P(v)$, is calculated as its frequency in ${C}_{i,j}$.

To reduce false positives, \ourmethod only uses extreme entropy patterns to refine the inferred semantic types.
The Static and Bytes types present two extremes of conveyed information: the Shannon entropy of the former is expected to be relatively small, and the latter to be large. \ourmethod then calculates the median Shannon entropy of all fields within the same cluster $C$. The inference result of the Static (Bytes) type is valid only when its Shannon entropy is smaller (larger) than the median.

\ourmethod also uses entropy characteristics to infer the semantic type of fields that lack execution information. It is based on the intuition that similar Shannon entropy usually appears in fields of the same semantic types.
For a field of unknown type, \ourmethod finds the field with the closest Shannon entropy within the same message cluster, and regards them as sharing the same type.

\begin{table}
  \centering \small
  \caption{Constraints between semantic functions and types.}
  \vspace{-1.0pc}
  \begin{tabular}{l|l}
    \hline
    \textbf{Field Function} & \textbf{Constraint on Field Type} \\
    \hline
    \textit{Command} & Should be \textit{Group}\\
    \textit{Length}& Should be \textit{Integer}\\
    \textit{Delim}& Should be \textit{Static} or \textit{Group}\\
    \textit{Aligned}& Should be \textit{Group} or \textit{Bytes}\\
    \textit{Checksum}& Should be \textit{Integer}\\
    \textit{Filename}& Should be \textit{String}\\
    \hline
  \end{tabular}
  \label{tab:type-function correlation}
  \vspace{-1.0pc}
\end{table}

\subsubsection{Type-Based Function Refinement.}
\label{sec:function_refinement}

We observe that there are some constraints between semantic types and functions. 
For a certain semantic function, its corresponding semantic type is constrained to a small subset. For example, the Length field should be the Integer type, as the data length is an integer value. 
Therefore, we utilize the refined semantic types to refine the semantic functions.
\ourmethod applies such constraints to remove the unreasonable inference results of semantic functions.
Table~\ref{tab:type-function correlation} summarizes the constraints.
The Command field should be of Group type, as the server typically selects one of the options for parsing an input message based on the Command value;
The Length field and Checksum field should be of Integer type according to their definitions;
The Delim field should be of Static or Group type, as it serves as data boundaries;
The Aligned field should be of Group or Bytes type, according to its typical implementations;
Similarly, the Filename field should be of String type.

\vspace*{2pt}
\noindent\textbf{Example-4.}
Figure~\ref{fig:illustrative_example4} illustrates the process of \emph{Semantic Refinement} module.
Assume it takes the given five messages as inputs. The left part shows the format and semantic results obtained from the \emph{Format Extraction} and \emph{Semantic Inference} modules. The right part shows the results obtained after executing the \emph{Semantic Refinement} module.
\ourmethod first iterates over all the message fields to identify the optimal clustering basis $f_7$ based on the similarity of message formats. It then figures out that the Shannon entropy of $f_{4,5}$ in the first cluster is zero, and thus removes the inference result of Bytes through entropy-based type refinement. Similarly, it also removes the unreasonable inference result Delim of $f_{4,5}$ through type-based function refinement.
With the \emph{Semantic Refinement module}, \ourmethod finally achieves the correct semantic inference of $f_{4,5}$.

\subsection{Implementation}
\label{sec:implementation}
\ourmethod is implemented in Python3 and C++.
The \emph{Execution Monitor} module taint tracks binary files with the dynamic instrumentation tool Pin~\cite{Pin}.
It uses Scapy~\cite{scapy} to derive execution information of messages, aiding subsequent field inference.
The \emph{Format Extraction} module implements instruction-based semantic similarity analysis from Section~\ref{sec:separator} to extract the message formats.
The \emph{Semantic Inference} module infers field semantics with the rules in Table~\ref{tab:types_funcs}, and validates filename conventions through file path syntax.
The \emph{Semantic Refinement} module follows the three-step process from Section~\ref{sec:corrector}, utilizing the results from \emph{Format Extraction} and \emph{Semantic Inference} modules to improve the overall correctness of field inference.
\ourmethod is modularly designed. The modules export uniform interfaces and could be easily extended to support additional field semantics.

\section{EVALUATION}

\label{sec:evaluation}

Our evaluation aims to answer the following research questions:
\begin{itemize}[leftmargin=*]
\item RQ1: How accurate is \ourmethod in performing format extraction?
\item RQ2: How accurate is \ourmethod in performing semantic inference?
\item RQ3: How effective are \ourmethod's semantic refinement components in improving the accuracy of semantic inference?
\item RQ4: How useful are the field inference results of \ourmethod in improving such downstream tasks as protocol fuzzing?
\end{itemize}

\subsection{Setup}

\subsubsection{Baselines.} \label{sec:baseline}
We compared \ourmethod with five state-of-the-art PRE tools:

\begin{itemize}[leftmargin=*] 
	\item \Polyglot~\cite{caballero2007polyglot} is an ExeT-based tool that infers key field semantics with simple heuristics to facilitate further format extraction.
	\item \AutoFormat~\cite{lin2008automatic} is an ExeT-based tool that extracts formats with hierarchical, parallel, and sequential relationships.
	\item \Tupni~\cite{cui2008tupni} is an ExeT-based tool that identifies field formats from instruction frequency and record sequences.
	\item \BinaryInferno~\cite{chandler2023binaryinferno} is a NetT-based tool that infers fields with multiple atomic detectors of different heuristic rules.
	\item \DynPRE~\cite{luodynpre} is a NetT-based tool that extracts field formats through the interactive capabilities of the server.
\end{itemize}

Among all baselines, \AutoFormat cannot infer semantics, while the others have very limited support. Since \Polyglot, \AutoFormat, and \Tupni are not publicly available, we re-implemented our versions of their techniques and validated with the examples in their papers before the comparison (detailed in Section~\ref{sec: Reproduction}).

\subsubsection{Benchmarks} 
We constructed the benchmarks as follows:

\noindent\textbf{Protocols}.
We selected 7 popular protocols evaluated by the prior PRE tools \Polyglot~\cite{caballero2007polyglot}, \AutoFormat~\cite{lin2008automatic}, \Tupni~\cite{cui2008tupni}, \BinaryInferno~\cite{chandler2023binaryinferno} and \DynPRE~\cite{luodynpre}.
These 7 protocols are widely-adopted and covering textual, binary, and mixed categories. Their application scenarios include industrial control and network communication. Among them, S7comm was a proprietary protocol, and the others are open-sourced.
We additionally selected another real-world protocol Ethernet/IP from prior work~\cite{shim2020clustering, wang2020ipart}. 
It is an industrial protocol that is widely adopted across many industrial sectors, including factory automation and hybrid process control.
Thus, we evaluated \ourmethod with 8 real-world protocols.
Among these 8 protocols, 7 have been evaluated by at least one prior PRE tool, and 4 have been evaluated by at least two prior tools. 
In particular, \ourmethod has 2, 1, 4, 2, 4 common protocols with \Polyglot, \AutoFormat, \Tupni, \BinaryInferno and \DynPRE, respectively.

\noindent\textbf{Messages}. Each protocol is tested with 50 messages with diverse and real-world usage scenarios. The messages are collected from two sources: (1) open-source network trace datasets; and (2) protocol clients or other relevant tools when the former is unavailable. 
Since the binary analysis based PRE techniques like \ourmethod are not sensitive to the sizes of input messages, we did not use a larger dataset size. 
On the other hand, the two network-based PRE techniques~(\BinaryInferno and \DynPRE) are designed to be capable of handling small-sized input messages.
Table~\ref{tab:information of datasets} in the Appendix lists the evaluated protocols and their characteristics including the server under testing, the message sources and the number of different message types.
For Ethernet/IP and HTTP, we ran the protocol client and the command line tool \textit{curl} with different options/parameters, respectively, to obtain diverse messages. 
For other protocols, we randomly filtered the input messages with different types and contents from their open-source traces to improve diversity and reduce dataset biases. 

\noindent\textbf{Ground-truth}. We adopted protocol specifications as the ground-truth. 
For each message, its ground-truth includes field boundaries (format) and field type/functions (semantics). Specifically, for each protocol's 50 messages, we use Wireshark's packet dissectors to parse each message and obtain the ground-truths of field formats and semantics. 
We also referred to the protocol RFCs for validation.

\vspace{2pt}

\subsubsection{Protocol fuzzing.} Protocol fuzzing is an important application scenario of PRE. To evaluate how well the PRE tools assist the downstream task performance, we incorporate their field inference results into a classical generation-based fuzzer, \boofuzz~\cite{boofuzz}, to guide test generation. 
Specifically, for the fields supported by PRE tools, we applied the corresponding types/functions in \boofuzz to these fields. For the unsupported fields, we adopted the \emph{random} strategy, which only leverages the format extraction results to limit the boundaries of each random field. Each scheme was tested for 10 hours on a machine with a Core i7-13700 CPU and 16GB memory.

\subsubsection{Metrics.} We adopt different metrics for different tasks.

\begin{figure}
  \centering
  \includegraphics[width=0.8\linewidth]{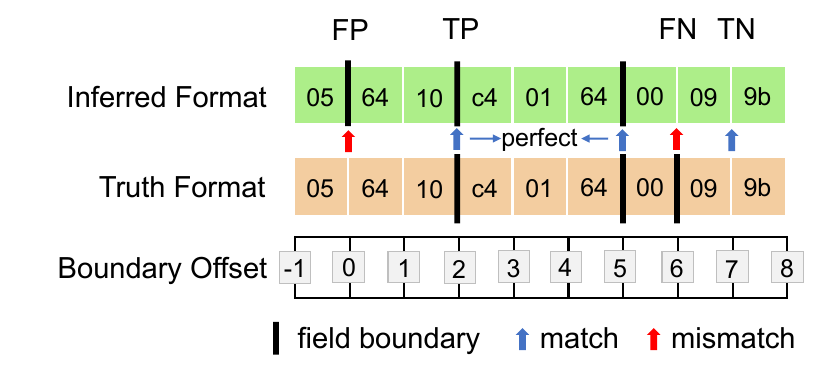}
  \vspace{-1.0pc}
  \caption{An illustration of format extraction metrics.}
  \label{fig:format metrix}
  \vspace{-0.1in}
\end{figure}

\textit{Format extraction} task is evaluated with the accuracy and F1-score~(a combination of precision and recall) of field boundary detection~\cite{chandler2023binaryinferno, luodynpre}.
As illustrated in Figure~\ref{fig:format metrix}, each boundary offset position between two adjacent bytes can fall into one of the four categories: FP, TP, FN, and TN.
Based on these definitions, we calculated \emph{accuracy}~(\ie, the number of correctly inferred positions out of all offset positions), \emph{precision}~(\ie, the number of inferred true field boundaries out of all inferred boundaries), and \emph{recall}~(\ie, the number of inferred true field boundaries out of all true boundaries).
We also counted the number of \textit{perfect} fields, of which both boundaries are accurately detected, and calculated \emph{perfection}~(\ie, the number of perfectly inferred fields out of all true fields).

\textit{Semantic inference} task is evaluated with the precision, recall, and F1-score~(a combination of precision and recall) of field semantic type/function identification. 
As different tools support different semantic types/functions~(see Section~\ref{sec:semantic inference}), we evaluated type/function separately. 
Specifically, \emph{precision} is calculated as the number of inferred true types/functions out of all inferred types/functions, and \emph{recall} is calculated as the number of inferred true types/functions out of all true types/functions.

\textit{Protocol fuzzing} is evaluated with branch coverage improvement. We used SanitizerCoverage~\cite{SanitizerCoverage} to identify covered unique branches. As any server's startup executes fixed program paths, we focus on the branches covered after the server has started.

\subsection{Results}

\begin{figure}
  \centering
  \includegraphics[width=0.7\linewidth]{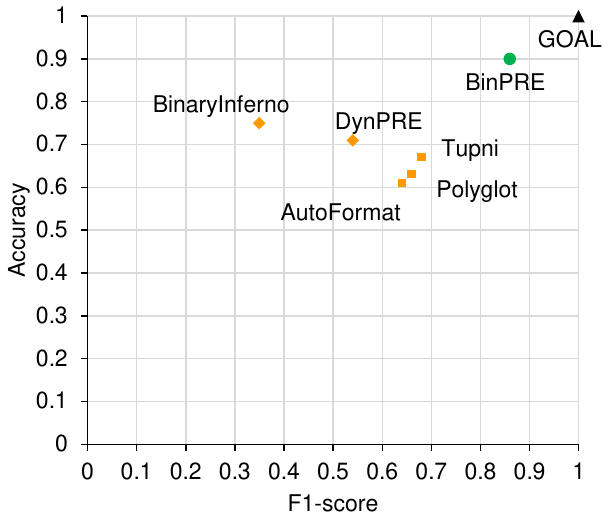}
  \vspace{-1.0pc}
  \caption{Average format extraction performance \small{(the more upper-right the point is, the better the tool)}}
  \label{fig:RQ1_pre_rec}
\vspace{-0.1in}
\end{figure}

\subsubsection{RQ1: Format.} Figure~\ref{fig:RQ1_pre_rec} summarizes the average performance of each tool on format extraction, and Table~\ref{tab:RQ1details} details the results.

Overall, \ourmethod significantly outperforms the baselines, achieving an average accuracy/F1-score/perfection of 0.90/0.86/0.73, due to its effective instruction-based semantic similarity analysis. In comparison, the baselines are 13-60\% lower. 

\begin{table*}
  \centering
  \caption{Format extraction result summary, including accuracy, F1-score, and perfection \small{(Bold numbers indicate the best results).} } 
  \vspace{-1.0pc}
  \setlength{\tabcolsep}{1.6mm}
  \footnotesize
  \begin{tabular}{c|ccc|ccc|ccc|ccc|ccc|ccc}
   \toprule
   \multirow{2}{*}{Protocol}  & \multicolumn{3}{c}{\ourmethod} & \multicolumn{3}{c}{\Polyglot} & \multicolumn{3}{c}{\AutoFormat} & \multicolumn{3}{c}{\Tupni} & \multicolumn{3}{c}{\BinaryInferno} & \multicolumn{3}{c}{\DynPRE}  \\
   \cmidrule(lr){2-4} \cmidrule(lr){5-7} \cmidrule(lr){8-10} \cmidrule(lr){11-13} \cmidrule(lr){14-16} \cmidrule(lr){17-19}
   & Acc.  & F1.   & Perf.   & Acc.  & F1.   & Perf.   & Acc.  & F1.   & Perf.   & Acc.  & F1.   & Perf.   & Acc.  & F1.   & Perf.   & Acc.  & F1.   & Perf.     \\
   \midrule
    Modbus  & \textbf{1.00}  & \textbf{1.00}  & \textbf{0.99}  & 0.91  & 0.93  & 0.84  & 0.91  & 0.93  & 0.84  & 0.82  & 0.87  & 0.61  & 0.79  & 0.79  & 0.32 & 0.71  & 0.79  & 0.39    \\ 
    S7comm  & 0.79  & 0.83  & 0.60  & 0.85  & 0.90  & \textbf{0.87}  & 0.82  & 0.88  & 0.80  & \textbf{0.86}  & \textbf{0.91}  & 0.85 & 0.54  & 0.52  & 0.09 & 0.65  & 0.71  & 0.28     \\ 
    Ethernet/IP  & \textbf{0.95}  & \textbf{0.92}  & 0.66  & 0.75  & 0.71  & \textbf{0.79}  & 0.52  & 0.56  & 0.38  & 0.72  & 0.69  & 0.75 & 0.77  & 0.57  & 0.28  & 0.90  & 0.83  & 0.42   \\ 
    DNP3.0  & \textbf{0.95}  & \textbf{0.95}  & \textbf{0.88}  & 0.90  & 0.84  & 0.63  & 0.65  & 0.72  & 0.75  & 0.75  & 0.67  & 0.25 & 0.61  & 0.50  & 0.25  & 0.59  & 0.36  & 0.08    \\ 
    DNS  & \textbf{0.74}  & \textbf{0.66}  & \textbf{0.62}  & 0.73  & \textbf{0.66}  & 0.58  & 0.39  & 0.46  & 0.44  & 0.63  & 0.58  & 0.47 & 0.82  & 0.46  & 0.11 & 0.45  & 0.41  & 0.26     \\ 
    FTP  & \textbf{0.88}  & \textbf{0.86}  & \textbf{0.57}  & 0.42  & 0.59  & 0.24  & 0.52  & 0.64  & \textbf{0.57}  & 0.56  & 0.63  & 0.07  & 0.72  & 0.00  & 0.00 & 0.79  & 0.67  & 0.36    \\ 
    TFTP  & \textbf{0.99}  & \textbf{0.96}  & \textbf{0.89}  & 0.25  & 0.36  & 0.30  & 0.44  & 0.43  & 0.30  & 0.83  & 0.74  & 0.30   & 0.89  & 0.00  & 0.00 & 0.77  & 0.46  & 0.07   \\ 
    HTTP  & \textbf{0.86}  & \textbf{0.72}  & 0.60  & 0.24  & 0.33  & 0.56  & 0.63  & 0.51  & \textbf{0.65}  & 0.20  & 0.32  & 0.58 & 0.83  & 0.00  & 0.00 & 0.82  & 0.14  & 0.03    \\ 
    \hline
   \textbf{\textit{Average}} & \textbf{0.90}  & \textbf{0.86}  & \textbf{0.73}  & 0.63  & 0.66  & 0.60  & 0.61  & 0.64  & 0.59  & 0.67  & 0.68  & 0.49 & 0.75  & 0.35  & 0.13 & 0.71  & 0.54  & 0.24   \\ 
   
   \bottomrule
  \end{tabular}
  \label{tab:RQ1details}
 \end{table*}

In some cases, \ourmethod has similar or slightly lower performance than the best baseline, due to the correlation between the format extraction algorithm and protocol characteristics. For example, the S7comm protocol has compact message structures and the behavioral differences between fields are relatively small (\ie several adjacent fields only retrieved by the parsing process).  Meanwhile, \Polyglot leverages the typical patterns of the single instruction-based field segmentation, which is particularly effective on such protocols. Therefore, it perfectly identifies 27\% more fields than \ourmethod. 
As another example, \ourmethod does not consistently achieve the highest perfection on the HTTP protocol, as the server uses delimiters as the starting point of the key-value pair field: the delimiters and their subsequent fields share similar semantics and are processed in similar ways. Thus, \ourmethod fails to identify their boundaries in some cases.

\begin{figure}[t]
  \centering
  \includegraphics[width=\linewidth]{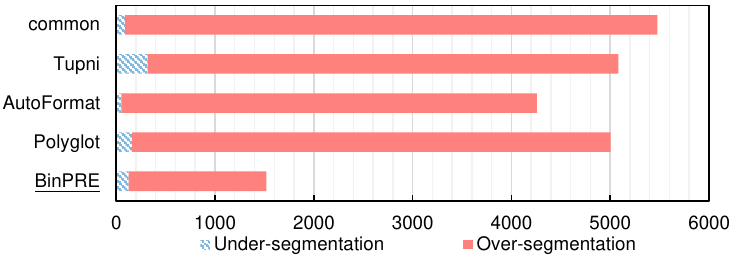}
  \vspace{-2.0pc}
  \caption{Under-segmentation and over-segmentation errors of the common strategy used by prior ExeT-based tools, the three ExeT-based baselines, and \ourmethod \small{(The segmentation errors caused by unused fields are excluded in this analysis).}}
  \label{fig:RQ1-unused}
  \vspace{-0.1in}
\end{figure}

Figure~\ref{fig:RQ1-unused} shows the numbers of over-segmentation and under-segmentation errors of the common strategy for format extraction used by the prior ExeT-based tools (detailed in Section~\ref{sec:format extraction}), the three ExeT-based baselines and \ourmethod, respectively.
We excluded the fields which have not been accessed by the servers because these fields cannot be inferred from the instruction traces.
Compared with the alternatives, \ourmethod has 63$\sim$70\% fewer segmentation errors, and achieves the best trade-off between over-segmentation and under-segmentation errors. While \AutoFormat has the least under-segmentation errors, it comes at the cost of more over-segmentation errors. 
Moreover, \ourmethod has 72\% fewer segmentation errors than the common strategy used by the prior ExeT-based tools.
\ourmethod's improvement is achieved by its effective instruction-based semantic similarity analysis strategy.

\vspace{-0.1in}
\begin{center}
\doublebox{\parbox{0.46\textwidth}{
    \emph{Answer to RQ1 (Format)}: \ourmethod identifies field formats with the perfection score of 0.73, which outperforms all the baselines.
}}
\end{center}

\begin{figure}[t]
  \centering
  \includegraphics[width=\linewidth]{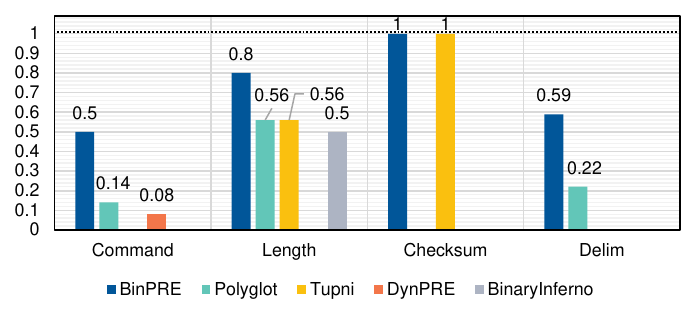}
  \vspace{-2.0pc}
  \caption{Average F1-scores of semantic inference across 8 protocols achieved by the evaluated PRE tools. The detailed results of each protocol/tool are given in Table~\ref{tab:RQ2details} in the Appendix \small{(Only the semantics supported by the tools are included)}.} 
  \label{fig:RQ2}
  \vspace{-0.1in}
\end{figure}

\subsubsection{RQ2: Semantics.} Overall, \ourmethod achieves an average precision, recall, and F1-score of 0.72/0.77/0.74 for semantic type inference and 0.77/0.91/0.81 for semantic function inference. 

We compared \ourmethod with the baselines, except for \AutoFormat which does not support semantics inference.
Figure~\ref{fig:RQ2} shows the average F1-score on each semantic which is supported by at least one baseline. 
\ourmethod greatly outperforms baselines in all cases. 
Specifically, \Polyglot and \Tupni are 24\% worse than \ourmethod on the Length field, as they heavily rely on behavior-based heuristic rules, which are not resilient towards diverse protocol implementations.  \DynPRE's and \BinaryInferno's  poor format extraction performance results in a high number of errors on semantic inference. Furthermore, the context-based rules of \BinaryInferno are not suitable for datasets of diverse formats.

\ourmethod also has relatively high performance on the other seven semantics that are unsupported by baselines: it has the F1-scores of 0.56 for Static, 0.61 for Integer, 0.50 for Group, 0.54 for Bytes, 0.15 for String, 0.00 for Aligned and 0.67 for Filename.
\ourmethod fails to identify the fields Aligned in our benchmark suite. Because the fields Aligned typically do not serve any functionality, and thus have little behavioral and contextual features.

\vspace{-0.1in}
\begin{center}
\doublebox{\parbox{0.46\textwidth}{
    \emph{Answer to RQ2 (Semantics)}: \ourmethod achieves the F1-scores of 0.74/0.81 on semantic inference for types/functions. It outperforms all the baselines on each semantic type/function.
}}
\end{center}

\subsubsection{RQ3: Ablation.} We conducted ablation studies to assess the effectiveness of \emph{Semantic Refinement} module in semantic inference.

\begin{table}
  \centering
  \footnotesize
  \caption{F1-score of semantic inference for field Command.}
  \vspace{-1.0pc}
  \setlength{\tabcolsep}{1.6mm}
  \begin{tabular}{|l|cc|cc|}
   \hline
   \textbf{Protocol} & \multicolumn{1}{c}{\ourmethoda} & \multicolumn{1}{c|}{\ourmethod} \\
   \hline
    Modbus  & 0.85  & 1.00   \\ 
    S7comm  & 0.22  & 1.00   \\ 
    Ethernet/IP  & 0.67  & 1.00   \\ 
    DNP3.0  & 0.00  & 0.00   \\ 
    TFTP  & 0.00  & 1.00   \\ 
    \hline
    \multirow{1}{*}{\textbf{Average}} & \multirow{1}{*}{0.35} & \multirow{1}{*}{0.80} \\
    \hline
    \end{tabular}
    \begin{tablenotes}
        \item * \ourmethoda is a variant of \ourmethod, which omits format-based clustering.
    \end{tablenotes}
  \label{tab:ablation study for clustering}
  \vspace{-0.1in}
\end{table}

\noindent\textbf{Format-based clustering.} Table~\ref{tab:ablation study for clustering} shows the Command field identification results with and without format-based clustering (\ourmethoda). As the format extraction module fails to correctly segment the command fields of DNS, FTP, and HTTP, we focus on the remaining protocols. The result shows that format-based clustering offers effective semantic refinement: \ourmethod achieves the F1-score of 0.8, while \ourmethoda drops to 0.35. Both \ourmethoda and \ourmethod fail on DNP3.0, as DNP's messages of different commands share similar formats, which conflicts with \ourmethod's conjunctures.

 \begin{table*}[ht]
  \centering
  \caption{\ourmethod's semantic inference results w/ and w/o type/function refinement \small{(Bold numbers indicates the best results)}.} 
  \vspace{-1.0pc}
  \setlength{\tabcolsep}{1.6mm}
  \footnotesize
  \begin{tabular}{l|ccc|ccc|ccc|ccc}
   \toprule
   \multirow{3}{*}{Protocol} & \multicolumn{6}{c}{\ourmethod Semantic Type} & \multicolumn{6}{c}{\ourmethod Semantic Function}\\
   & \multicolumn{3}{c}{w/o Refinement} & \multicolumn{3}{c}{w/ Refinement} & \multicolumn{3}{c}{w/o Refinement} & \multicolumn{3}{c}{w/ Refinement}\\
   \cmidrule(lr){2-4} \cmidrule(lr){5-7} \cmidrule(lr){8-10} \cmidrule(lr){11-13}
   & Pre.  & Rec.   & F1.   & Pre.  & Rec.   & F1.   & Pre.  & Rec.   & F1.   & Pre.  & Rec.   & F1.    \\
   \midrule
    Modbus  & \textbf{0.83}  & 0.61  & 0.70  & \textbf{0.83}  & \textbf{0.84}  & \textbf{0.83}  & 0.43  & 0.87  & 0.57  & \textbf{0.77}  & \textbf{1.00}  & \textbf{0.87}  \\ 
    S7comm  & 0.50  & 0.53  & 0.52  & \textbf{0.51}  & \textbf{0.55} & \textbf{0.53}  & 0.42  & \textbf{0.56}  & 0.48  & \textbf{0.85}  & \textbf{0.56}  & \textbf{0.67}   \\ 
    Ethernet/IP  & 0.60  & 0.77  & 0.67  & \textbf{0.80}  & \textbf{0.90}  & \textbf{0.85}  & 0.46  & 1.00  & 0.63  & \textbf{1.00}  & \textbf{1.00}  & \textbf{1.00}   \\ 
    DNP3.0  & 0.55  & \textbf{0.70}  & 0.62  & \textbf{0.71}  & \textbf{0.70}  & \textbf{0.71}  & 0.43  & \textbf{1.00}  & 0.60  & \textbf{0.77}  & \textbf{1.00}  & \textbf{0.87}   \\ 
    DNS  & \textbf{0.59}  & \textbf{0.79}  & \textbf{0.67}  & \textbf{0.59}  & \textbf{0.79}  & \textbf{0.67}  & 0.25  & \textbf{0.98}  & 0.40  & \textbf{0.29}  & \textbf{0.98}  & \textbf{0.45}   \\ 
    FTP  & \textbf{0.62}  & \textbf{0.62}  & \textbf{0.62}  & \textbf{0.62} & \textbf{0.62} & \textbf{0.62}  & 0.87  & \textbf{1.00}  & 0.93  & \textbf{0.90}  & \textbf{1.00}  & \textbf{0.95}   \\ 
    TFTP  & 0.50  & 0.67  & 0.57  & \textbf{0.75}  & \textbf{1.00}  & \textbf{0.86}  & 0.33  & 0.50  & 0.40  & \textbf{0.67}  & \textbf{1.00}  & \textbf{0.80}   \\ 
    HTTP  &\textbf{0.95} & \textbf{0.92}  & \textbf{0.94}  & 0.94  & 0.78  & 0.85  & 0.94  & \textbf{0.93}  & \textbf{0.93}  & \textbf{0.95}  & 0.77  & 0.86   \\ 
   \hline
   \textbf{\textit{Average}} & 0.64  & 0.70  & 0.66  & \textbf{0.72}  & \textbf{0.77}  & \textbf{0.74}  & 0.52  & 0.85  & 0.62  & \textbf{0.77}  & \textbf{0.91}  & \textbf{0.81}  \\ \hline
   \bottomrule
  \end{tabular}
  \label{tab:RQ3}
 \end{table*}

\noindent\textbf{Type/function refinement.} We compared \ourmethod with its variant which omits type/function refinement. As shown in Table~\ref{tab:RQ3}, adopting type~(function) refinement improves the semantic inference's precision by 0.08~(0.25), recall by 0.07~(0.06), and F1-score by 0.08~(0.19). It is worthy noting that adopting both refinements improves the F1-score of Command by 0.29, Length by 0.26, Bytes by 0.18, and Group by 0.15.
These improvements are achieved by handling the complex and irregular protocol implementations by refining inference results with extra contextual information.

\vspace{-0.1in}
\begin{center}
\doublebox{\parbox{0.46\textwidth}{
    \emph{Answer to RQ3 (Ablation)}: All the three components of \emph{Semantic Refinement} module are effective in refining semantic inference results. \emph{Format-based clustering} improves the F1-score of command inference from 0.35 to 0.80.
    \emph{Type(function) refinement} improves the F1-score of semantic results from 0.66(0.62) to 0.74(0.81).
}}
\end{center}

\begin{table*}
  \centering
  \caption{Speedup of BinPRE-enhanced \boofuzz variants (\ie, \ourmethod, \ourmethodc and \ourmethodd) over those enhanced by prior PRE tools in achieving highest identical branch coverage \small{(Numbers larger than 1x indicate higher speeds of achieving coverage)}.}
  \vspace{-1.0pc}
  \setlength{\tabcolsep}{1.6mm}
  \footnotesize
  \begin{tabular}{c|ccc|ccc|ccc|ccc}
   \toprule
   \multirow{2}{*}{}  & \multicolumn{3}{c}{Modbus} & \multicolumn{3}{c}{Ethernet/IP} & \multicolumn{3}{c}{HTTP} & \multicolumn{3}{c}{FTP}  \\
   \cmidrule(lr){2-4} \cmidrule(lr){5-7} \cmidrule(lr){8-10} \cmidrule(lr){11-13} 
   & \ourmethod  & \ourmethodc   & \ourmethodd   & \ourmethod  & \ourmethodc   & \ourmethodd   & \ourmethod  & \ourmethodc   & \ourmethodd   & \ourmethod  & \ourmethodc & \ourmethodd  \\
   \midrule
    Polyglot & 408.7x & 670.7x & 0.2x & 513.3x & 365.8x & 154.8x & 0.5x & 2.7x & 3.9x & 0.7x & 3.1x & 5.4x  \\ 
    Tupni & 975.1x & 1628.8x & 0.4x & 20.1x & 14.5x & 2.1x & 5.1x & 1.1x & 1.7x & 0.8x & 2.5x & 6.7x  \\ 
    AutoFormat & 3585.9x & 5986.5x & 0.5x & 49.9x & 35.9x & 3.7x & 26.9x & 4.7x & 8.6x & 0.7x & 4.7x & 38.9x  \\ 
    BinaryInferno & 82.6x & 76.0x & 45.1x & 6989.0x & 4997.4x & 1752.6x & 3972.0x & 4038.4x & 2383.0x & 84.0x & 70.2x & 101.1x  \\ 
    DynPRE & 5.5x & 4.6x & 2.7x & 1134.1x & 807.5x & 0.0006x & 1.4x & 0.7x & 1.0x & 1233.6x & 1761.9x & 1606.7x  \\ 
    Random & 29.5x & 29.4x & 35.2x & 8.7x & 6.1x & 3.9x & 36.1x & 36.7x & 12.8x & 133.6x & 111.6x & 160.7x \\ \hline
  \end{tabular}
  \label{tab:RQ4speed}
 \end{table*}

\subsubsection{RQ4: Downstream tasks.} 
\begin{figure}[t]
	\centering
	\includegraphics[width=\linewidth]{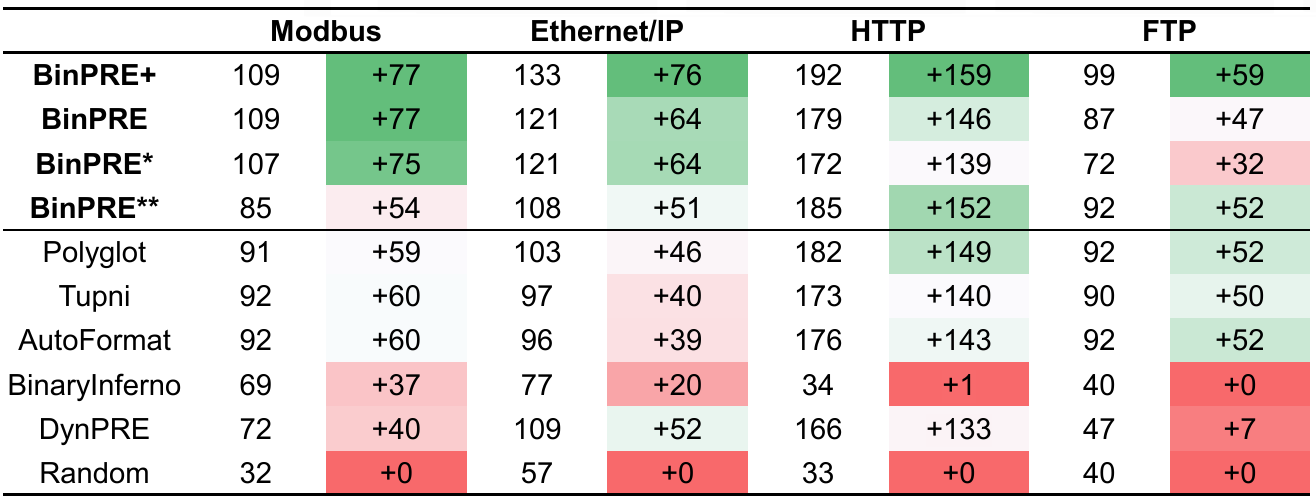}
	\vspace{-2.0pc}
	\caption{Numbers of unique branches covered by PRE-enhanced \boofuzz (averaged across three repeated runs).}
\label{fig:RQ4-2}
\vspace{-0.1in}
\end{figure}

\begin{figure*}[t]
	\centering
	\includegraphics[width=0.9\linewidth]{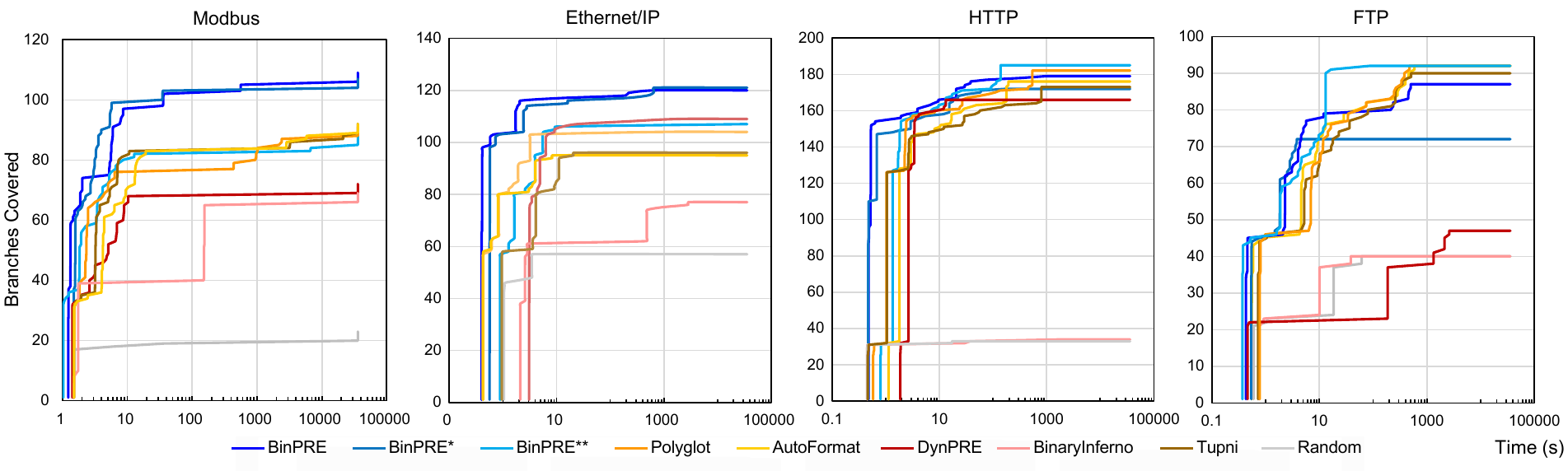}
	\vspace{-1.0pc}
	\caption{Numbers of unique branches covered by \boofuzz in ten hours enhanced by the results of different PRE tools.}
	\label{fig:RQ4-1}
\end{figure*}

We evaluated the field inference results from \ourmethod and the baselines to enhance the downstream task of protocol fuzzing.
Since \ourmethod provides the information of both field formats and semantics (types and functions), we also evaluated the two variants of \ourmethod: (1) \ourmethodc (only keeping the field formats and types) and (2) \ourmethodd (only keeping the field formats). 
Figures \ref{fig:RQ4-2} and ~\ref{fig:RQ4-1} summarize the results of branch coverage by fuzzing four representative protocols, \ie, Modbus, Ethernet/IP, HTTP and FTP, from our benchmark suite. Modbus and Ethernet/IP are binary protocols, while HTTP and FTP are text protocols. 
To mitigate the randomness during fuzzing, we ran each configuration for three times and computed the average values.

\noindent\textbf{Branch Coverage}.  
In Figure~\ref{fig:RQ4-2}, \ourmethode denotes the union results of unique branches covered by \ourmethod, \ourmethodc and \ourmethodd. It indicates the overall usefulness of the PRE results (with different levels of information) of \ourmethod to enhance \boofuzz.
The Random strategy denotes the results of the vanilla \boofuzz without given any results of the PRE tools (\ie, \boofuzz simply generate random message contents).
In Figure~\ref{fig:RQ4-2}, for each protocol, we give the numbers of branches covered by \boofuzz enhanced by the results of different PRE tools, and the improved branch coverage \wrt the random strategy.
On average, \ourmethod-enhanced Boofuzz (denoted by \ourmethode) achieves 20\%/29\%/5\%/8\% (51\%/22\%/16\%/111\%) higher branch coverage on Modbus/Ethernet IP/HTTP/FTP, compared to the best ExeT-based (NetT-based) baseline \Polyglot (\DynPRE). \ourmethod-enhanced Boofuzz improves the performance of vanilla \boofuzz~(denoted by Random)  by covering 244\%/133\%/482\%/148\% more branches on Modbus/Ethernet IP/HTTP/FTP. 
\ourmethod is particularly effective in improving fuzz testing on binary protocols because the binary protocols have compact and diverse field formats and semantics. It covers 244\% more branches on Modbus, and 133\% more on Ethernet/IP. 
On the loosely formatted text protocols, the information of field formats is more useful for fuzzing: \ourmethodd covers more branches on HTTP/FTP than \ourmethod/\ourmethodc.

\noindent\textbf{Efficiency}
Figure~\ref{fig:RQ4-1} shows the numbers of unique branches covered by \boofuzz enhanced by the results of different PRE tools. 
We can see that \ourmethod-enhanced \boofuzz variants~(\ie, \ourmethod, \ourmethodc and \ourmethodd) outperforms those enhanced by the prior PRE tools in most cases.
In Table~\ref{tab:RQ4speed}, we computed the speedup of \ourmethod-enhanced \boofuzz variants over those enhanced by the prior PRE tools in achieving highest identical branch coverage.
The numbers greater than 1.0x indicate higher speeds of achieving branch coverage.
We can see that \emph{in most cases} \ourmethod-enhanced \boofuzz variants are faster than the other baselines.

\noindent\textbf{Revealed CVEs.}  
\ourmethod helped \boofuzz discover a new zero-day vulnerability CVE-2024-31504 (CVSS score of 7.5 HIGH)
in FreeMODBUS.
It could cause a buffer overflow and thus lead to a denial of service. 
This vulnerability is related to two critical fields: (1) a length field (in the message's header) and (2) a variable-length field (in the message's payload). It is triggered when the value of the former is large enough to allow the latter to exceed the buffer limit, which causes a buffer overflow when FreeMODBUS tries to read it.
Specifically, a specific protocol variable \textit{n} is overwritten by a variable-length field, which indexes the array \textit{fd\_bits}, causing a segmentation fault when \textit{n} is overwritten to a large integer.

Only \ourmethod helps \boofuzz discover this CVE, as \ourmethod is the only PRE tool that correctly segments the two critical fields and correctly infers their types/functions --- a Length field with integer type, and a field with Bytes type. 
During the fuzzing, the inferred Length function renders the server to process the entire message payload, and the inferred Bytes type facilitates the generation of byte sequences with arbitrary length. 
However, all the baseline tools fail to identify these two fields.  AutoFormat, Tupni, BinaryInferno, and DynPRE fail in format extraction, while the other tools fail in semantic inference. 
Therefore, the baselines cannot find this CVE.
Additionally, \ourmethod also helps \boofuzz find one CVE-requested bug (a buffer overflow vulnerability in FreeMODBUS) and one known vulnerability CVE-2020-29596 in Miniweb.

\begin{center}
\vspace{-0.2in}
\doublebox{\parbox{0.46\textwidth}{
    \emph{Answer to RQ4 (Downstream tasks)}: \ourmethod can enhance protocol fuzzing, and help discover new or known vulnerabilities. The \ourmethod-enhanced \boofuzz achieves 5$\sim$29\% higher branch coverage, compared to the best prior ExeT-based PRE tool. 
}}
\end{center}
\vspace{-0.1in}

\vspace{-0.3pc}
\subsection{Discussions}

\noindent\textbf{Cost of \ourmethod.}
The runtime cost of \ourmethod includes two parts: (1) tracking execution information of input messages; and (2) inferring format and semantics from the execution information. 
For the first part, \ourmethod's instrumentation introduces average 59\% runtime overhead. For the second part, analyzing 50 messages takes less than 2 minutes on a machine with a Core i5-1038NG7 CPU and 8GB memory.
As protocol reverse engineering is typically a one-time cost, we believe that such overhead is affordable, considering \ourmethod's benefits on downstream tasks. 

\noindent\textbf{Reproduction of Baselines.}
\label{sec: Reproduction}
Three of the baselines (\Polyglot, \AutoFormat and \Tupni) are not publicly available. Therefore, we re-implemented them according to their algorithm descriptions and validated our implementations with the given examples in their original papers. We observe that our evaluation results are also consistent with their claims.

\begin{itemize}[leftmargin=*]
    \item \Polyglot: Sections 3--6 of its paper were re-implemented. Its format extraction algorithm and direction field identification strategy are validated on the DNS protocol. Its keyword and separator~(delimiter) identification strategies are validated on the HTTP protocol. Our implementation achieved the same results as described in the paper. 
    Note that Command and Length are the most common keyword and direction fields, respectively~\cite{caballero2007polyglot}. 
    Therefore, their identification strategies were implemented and compared in our evaluation. 
    \item \AutoFormat: Section 3 (excluding 3.2.2) of its paper was re-implemented. Its format extraction algorithm is validated on the HTTP protocol, achieving the same results as in the paper.
    \item \Tupni: Sections 3.1--3.5 of its paper were re-implemented. Its format extraction algorithm is validated on the HTTP protocol. Our implementation has slightly different outputs from the paper. These discrepancies are caused by the unused fields and delimiter checking within the server, which is consistent with the root causes \ie, ``a program ignores certain parts of an input''. Note that neither the original \Tupni nor our implementation supports delimiter recognition. Therefore, we believe that our implementation could reflect the original capability of \Tupni.
\end{itemize}

\vspace{2pt}
The other baselines are open-sourced and we directly used them.
While their performances align with their paper descriptions, there are variations on \DynPRE.
For instance, \DynPRE performed worse on the HTTP protocol in our evaluation because it segments fields based on dynamic analysis of server responses, limiting context information when a protocol generates similar responses for different inputs. This issue was confirmed by \DynPRE's authors.

\noindent\textbf{Effect of Data Size}.
We also evaluated the effectiveness of \ourmethod on small-sized input messages. We created a 5-message dataset by randomly sampling from the 50-message one used in our evaluation. \ourmethod has achieved similar performance on the 5-message dataset with the average 
F1-scores of 0.87 for format extraction and 0.75/0.84 for semantic type/function inference, respectively.
It shows that \ourmethod does not rely on large-sized input messages and could maintain the similar performance on small-sized ones. 

\noindent\textbf{Limitations}.
The taint analysis module of \ourmethod only works at byte-level. It cannot accurately deal with bit-level fields like bit-flags. 
To our knowledge, all existing PRE tools take byte as the basic unit for field inference due to the dominance of byte-level fields. 

\section{RELATED WORK}

\label{sec:related work}

\subsection{Automated Protocol Reverse Engineering}

Existing work proposes two types of PRE techniques: (i) network traffic-based (NetT-based) inference; and (ii) execution trace-based (ExeT-based) inference~\cite{huang2022protocol}.
NetT-based inference~\cite{cui2007discoverer, wang2012semantics,bossert2014towards,chen2021inspector,bermudez2016towards, ye2021netplier,pohl2019automatic,kleber2018nemesys} leverages numerous messages within sessions and investigates their relationships.  It infers protocol specifications by capturing message features from sessions\jiang{i removed ''real-world''}.
ExeT-based inference~\cite{caballero2007polyglot, lin2008automatic, cui2008tupni, comparetti2009prospex, wang2009reformat, caballero2009dispatcher, wang2008towards, lin2010automatic,cho2010inference, liu2013extracting} takes the binary files of servers as inputs.  It monitors the execution trace and tracks the message parsing process in binary files, which reveals the internal design logic.
With these execution information, it segment messages into fields and infers field semantics. 
While \ourmethod and the prior ExeT-based work all use taint analysis, they have distinct differences in format extraction, semantic inference and refinement. 
We will discuss the differences in the next two sections.

\subsection{PRE for Format Extraction}
{Format extraction} is an essential PRE task.
NetT-based approaches identify field boundaries through statistical methods.
\textsc{Netzob}~\cite{bossert2014towards} and \textsc{Netplier}~\cite{ye2021netplier} adopt alignment-based methods.
\textsc{Discoverer}~\cite{cui2007discoverer} recursively clusters messages of the same type.
\BinaryInferno~\cite{chandler2023binaryinferno} treats messages as information sequences and proposes an information theoretic approach method.
\DynPRE~\cite{luodynpre} further utilizes response messages to enhance boundary identification.
These approaches rely on diverse large-scale message data.

ExeT-based approaches monitor the execution traces to extract field boundaries.
\Polyglot~\cite{caballero2007polyglot} is one of the first efforts to employ taint analysis techniques to monitor the binary files of target protocols and capture execution traces.
It uses three specific field functions~(\textit{direction field}, \textit{keyword}, and \textit{separator}), which are inferred by heuristic patterns of execution traces, to segment fields.
Building upon the format partitioning framework of \Polyglot, \AutoFormat~\cite{lin2008automatic} and \Tupni~\cite{cui2008tupni} propose hierarchical and packet field recognition methods. 
The former segment fields are based on the locality of executed instructions and the similarity of their call stacks.
The latter identifies candidate fields based on byte occurrences and merges the consecutive bytes processed in one loop into one field.
\textsc{Prospex}~\cite{comparetti2009prospex} further extends \AutoFormat to the session level, tracking and dividing the message parsing process within a complete session based on system calls.
These techniques rely on the behavioral features observed in execution traces of message parsing. 
However, due to the variations in protocol implementations, these features may not reflect the actual message formats. 
To address this issue, \ourmethod compares the semantic similarity by approximating the operator sequences between executed instruction sequences.
The instruction-based semantic similarity analysis of \ourmethod can capture deeper internal relationship between bytes and is more resilient to different protocol implementations.

\subsection{PRE for Semantic Inference}

{Semantic inference} focuses on understanding the meanings of each field in the messages~\cite{caballero2013automatic}. 
NetT-based approaches face obstacles to inferring protocol semantics, as the network traces only contain syntax information~\cite{2007Rosetta}. While researchers propose clustering~\cite{krueger2010asap, beddoe2004network, bermudez2015automatic}, supervised deep-learning~\cite{yang2020deep, zhao2022prosegdl} and other heuristic~\cite{markovitz2017field, ladi2018message} solutions to additionally support semantic inference, they highly rely on large-scale diverse network packets. In addition, they are only able to handle limited field semantics and have low inference accuracy~(illustrated in Table~\ref{tab:Qualitative Analysis}). 

ExeT-based inference approaches extract field semantics from the execution information of the protocol binaries.
One line of work, including \textsc{Dispatcher}~\cite{caballero2009dispatcher}, captures library function semantics to infer field semantics. 
Another line of work~\cite{caballero2007polyglot, cui2008tupni}, designs heuristic rules to analyze the execution traces of fields and identify behavioral semantic features, which are then mapped to field semantics.
However, these work suffer from inadequate and inaccurate semantic inference due to their limited and inaccurate behavior based rules.
For instance, \textsc{Polyglot} and \textsc{Tupni} infer only three and two semantic functions with limited accuracy, respectively.

\ourmethod differs from them in two aspects: 1) we are the first to construct a library of atomic semantic detectors for semantic inference; and 2) \ourmethod utilizes contextual information to refine the semantic inference results in a systematic way.
The former addresses the inadequate rules of semantic inference, while the latter mitigates the inaccurate results of semantic inference.

\subsection{Protocol Fuzzing}
Protocol fuzzing is widely used for testing protocol implementations, and generates massive test cases to trigger abnormal runtime behaviors~\cite{zhang2024survey, 9482063}. 

Mutation-based fuzzers (\eg, \textsc{NSFuzz}~\cite{qin2023nsfuzz}, \textsc{AFLNet}~\cite{pham2020aflnet}, and \textsc{ChatAFL}~\cite{meng2024large}) require initial seeds with proper formatting and content. They then employ various mutation methods on these message seeds. 
Generation-based fuzzers (\eg, \boofuzz~\cite{boofuzz} and \peach~\cite{peach}) generate semi-valid messages based on the given templates. Their effectiveness heavily depends on the pre-defined protocol structures. 
However, real-world protocols typically have complex structures and communication logic. Manually constructing initial seeds and templates requires huge efforts and also knowledge of the protocol specifications. 
\ourmethod tackles a different problem from these work. It provides automated PRE solutions to assist protocol fuzzing.

\section{CONCLUSION}

\label{sec:conclusion}
Protocol reverse engineering aims to infer the specifications of closed-source protocols. 
In this paper, we propose \ourmethod, a binary analysis based PRE tool that supports both format extraction and semantic inference.
It incorporates an instruction-based semantic similarity analysis for format extraction, and a novel library composed of atomic semantic detectors and a cluster-and-refine paradigm for improving semantic inference.
We evaluate \ourmethod with a variety of protocols and achieve high accuracy on the format extraction and semantic inference tasks. It also effectively supports the downstream task of protocol fuzzing.

\bibliographystyle{ACM-Reference-Format}
\balance
\bibliography{references}

\clearpage
\appendix

\section*{Appendix.A \quad Semantic Types and Functions in \boofuzz and \peach}

\label{sec:appendix A}
\subsection*{A.1\quad Semantics supported by \boofuzz}

\boofuzz supports all the semantics summarized in Table~\ref{tab:Qualitative Analysis}, which includes five types and six functions.

The five types are:
\begin{itemize}
    \item Integer: it represents a number of variable length. It is an abstraction of \texttt{BitField}, \texttt{Byte}, \texttt{Word}, \texttt{DWord}, and \texttt{QWord} in \boofuzz.
    \item Static: it is fixed and does not vary with specific content, which is equivalent to \texttt{Static} in \boofuzz.
    \item Group: it comprises a list of available values, which is equivalent to \texttt{Group} in \boofuzz.
    \item Bytes: it denotes a sequence of binary bytes of arbitrary length, which is equivalent to \texttt{Bytes} in \boofuzz.
    \item String: it denotes a sequence of characters of arbitrary length, which is equivalent to \texttt{String} in \boofuzz.
\end{itemize}

The six functions are:
\begin{itemize}
    \item Command: it denotes the message type, which distinguishes the format of messages with different types in \boofuzz.
    \item Length: it records the size of the message or data, which is equivalent to \texttt{Size} in \boofuzz.
    \item Filename: it refers to a file within the file system, which is equivalent to \texttt{Fromfile} in \boofuzz.
    \item Delim: it indicates the end of a text protocol field, which is equivalent to \texttt{Delim} in \boofuzz.
    \item Checksum: it verifies the integrity of messages or data, which is equivalent to \texttt{Checksum} in \boofuzz.
    \item Aligned: it aligns data, which is equivalent to \texttt{Aligned} in \boofuzz.
\end{itemize}

\subsection*{A.2\quad Semantics supported by \peach}

\peach supports three types and four functions in  Table~\ref{tab:Qualitative Analysis}.

The three types are:

\begin{itemize}
    \item Integer: \peach has \texttt{Flag}/\texttt{Flags} field, which defines a specific bit field, and  \texttt{Number} that defines a binary number of the specified length. They serve as Integer fields. 
    \item Group: it refers to the \texttt{Choice} field.
    \item String: it refers to the \texttt{String} field.
\end{itemize}

The four functions are:

\begin{itemize}
    \item Command: it denotes the message type, which distinguishes the \texttt{DataModel} in \peach.
    \item Length: the \emph{Size-of Relation} in \texttt{Relation} field is consistent with the Length function. 
    \item Checksum: the \emph{Checksum Fixups} in \texttt{Fixup} field provides the Checksum function. 
    \item Aligned: it refers to the \texttt{Padding} field in \peach.
\end{itemize}

Note that, \peach also supports two additional functions: the \texttt{Placement} function determines the movement of specific elements after parsing the input stream; and the \texttt{Transformer} function  performs static transformations or encoding on the parent element.

\section*{Appendix.B \quad Extending prior work with our library of atomic semantic detectors}
\label{sec:appendix BB}

It is fundamentally impossible for prior work to achieve similar semantic inference results if we extend them with our own library of atomic semantic detectors. Because the semantic inference relies on the quality of format extraction.
To this end, we did an additional experiment by integrating our own library of atomic semantic detectors~(Section~\ref{sec:speculator}) into Polyglot, AutoFormat, and Tunpi.
The evaluation results are shown in Table~\ref{tab:library_enhanced_details}.
On average,  Polyglot, AutoFormat and Tunpi only achieved the F1-scores of 0.34, 0.28, and 0.25 for type inference, and 0.22, 0.23, and 0.22 for function inference, respectively. 
In comparison, \ourmethod achieved the F1-scores of 0.53~(0.46) for type inference, and 0.64~(0.45) for function inference w/~(w/o) semantic refinement, respectively.

\section*{Appendix.C \quad Justifying the similarity threshold used in the Needleman-Wunsch~(NW) algorithm }
\label{sec:appendix CC}

The standard convention often sets the similarity in the Needleman-Wunsch~(NW) algorithm at 0.8. 
To further study the effectiveness of this threshold on \ourmethod, we explored  the 0-1 interval in 0.1 steps to identify the threshold range where \ourmethod yielded the most optimal results in evaluating each protocol.
For DNS, TFTP, DNP3.0, Modbus, and FTP, the optimal threshold range is 0.1-1.0.
For S7comm and Ethernet/IP, the optimal threshold ranges are 0.5-1.0 and 0.6-1.0, respectively.
For HTTP, the optimal threshold range is 0.1-0.4.
Since 0.8 is within the optimal threshold ranges of most protocols, it is appropriate for our method to set the NW threshold at 0.8.

\section*{Appendix.D \quad Results of format extraction of the classic PRE techniques.}

\begin{figure*}[hb]
  \centering
  \includegraphics[width=0.5\linewidth]{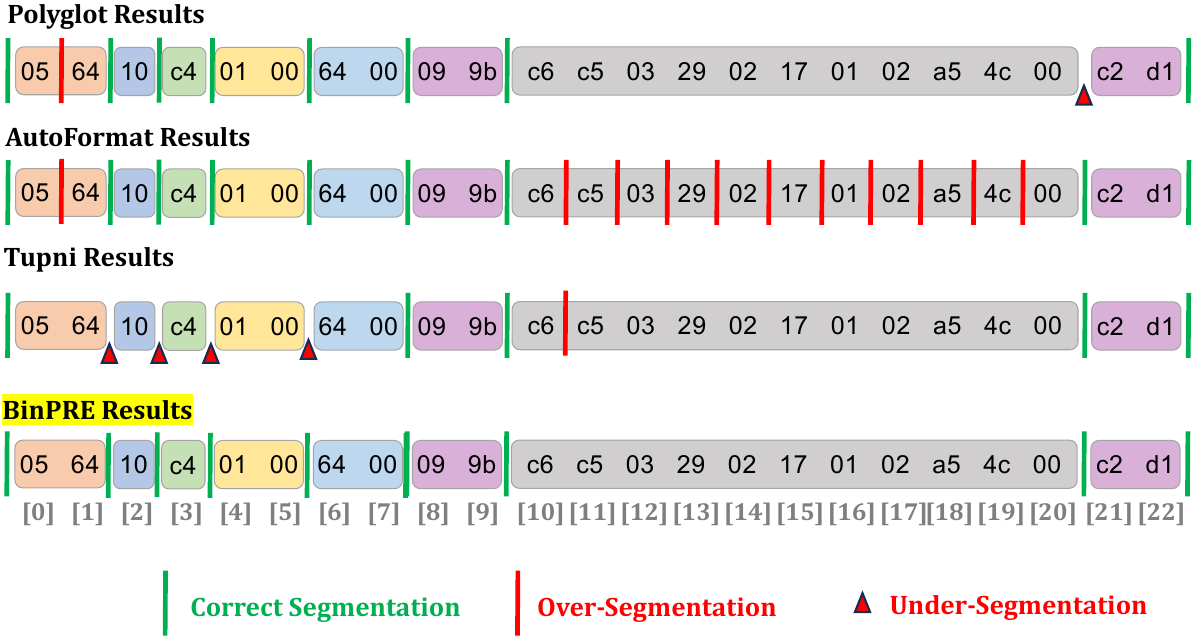}
  \vspace{-1.0pc}
  \caption{The format extraction results of classic PRE techniques on the example message~(the fields with different semantics are annotated by different colors).}
  \label{fig:dnp3_example_baselines}
\end{figure*}

Figure~\ref{fig:dnp3_example_baselines} illustrates the format extraction results of three classic PRE techniques and \ourmethod when applied to process the message in Figure~\ref{fig:dnp3_example_new}.

\Polyglot, \AutoFormat, and \Tupni all suffer from the over-segmentation errors when the bytes belonging to one field are accessed by different instructions, as they follow the common strategy~(detailed in Section~\ref{sec:format extraction}).
Despite incorporating their own heuristic strategies to mitigate over-segmentation errors, these strategies may unintentionally lead to under-segmentation errors.

Comparatively, \ourmethod introduces an instruction-based semantic similarity analysis strategy for message format extraction. This strategy mitigates the errors of both over-segmentation and under-segmentation, and accurately segments the protocol messages.

\begin{table*}[hb]
  \centering
  \caption{F1-scores of semantic inference of the prior PRE tools by extending them with our library of atomic semantic detectors. } 
  \vspace{-1.0pc}
  \setlength{\tabcolsep}{1.6mm}
  \footnotesize
  \begin{tabular}{c|ccccc|ccccc|}
   \toprule
   \multirow{2}{*}{Protocol}  & \multicolumn{5}{c}{Semantic Type} & \multicolumn{5}{c}{Semantic Function} \\
   \cmidrule(lr){2-6} \cmidrule(lr){7-11} 
   & \ourmethod  & \ourmethodb   & \Polyglota   & \AutoFormata  & \Tupnia   & \ourmethod  & \ourmethodb   & \Polyglota   & \AutoFormata  & \Tupnia        \\
   \midrule
    Modbus  &  0.80 & 0.70  & 0.68  & 0.68  & 0.52  &  0.86 & 0.59  & 0.38  & 0.38  &  0.34     \\ 
    S7comm  & 0.36  & 0.38  & 0.47  &  0.45 & 0.47  & 0.55  &  0.44 &  0.23 &  0.23 & 0.23     \\ 
    Ethernet/IP  & 0.63  & 0.53  &  0.48 & 0.19  & 0.45  & 1.00  & 0.54  & 0.24  & 0.23  & 0.23    \\ 
    DNP3.0  & 0.55  & 0.49  & 0.42  & 0.31  & 0.00  & 0.76  & 0.55  & 0.40  & 0.36  & 0.44      \\ 
    DNS  & 0.48  & 0.50  & 0.47  & 0.24  & 0.37  & 0.27  & 0.24  & 0.23  & 0.18  & 0.27       \\ 
    FTP  & 0.27  & 0.28  & 0.00  & 0.19  & 0.00  & 0.55  & 0.53  & 0.09  & 0.30  & 0.00       \\ 
    TFTP  & 0.79  &  0.54 &  0.00 & 0.00  &  0.00 & 0.80  & 0.40  & 0.00  &  0.00 &  0.00     \\ 
    HTTP  & 0.35  & 0.28  & 0.18  & 0.18  & 0.18  & 0.35  & 0.34  & 0.18  & 0.17  & 0.21       \\ 
    \hline
   \textbf{\textit{Average-50}} & 0.53  & 0.46  & 0.34  & 0.28  &  0.25 & 0.64  & 0.45  & 0.22  & 0.23  &  0.22  \\ 
   
   \bottomrule
  \end{tabular}
      \begin{tablenotes}
        \item * \ourmethodb is a variant of \ourmethod, which omits semantic refinement module.
        \item * \Polyglota is a variant of \Polyglot, which is extended to include our library.
        \item * \AutoFormata is a variant of \AutoFormat, which is extended to include our library.
        \item * \Tupnia is a variant of \Tupni, which is extended to include our library.
        \end{tablenotes}
  \label{tab:library_enhanced_details}
 \end{table*}

\begin{table*}[hb]
  \centering
  \footnotesize
  \caption{Protocols and messages of our benchmark suite.}
  \vspace{-1.0pc}
    \begin{tabular}{cccccccccccccccccccccc}
    \hline
    Protocol & Content & Scenario & Project & Server Under Testing & \multicolumn{2}{c}{Message Source \& \# of Message Types} \\
    \hline
    Modbus & binary & control & freemodbus\cite{freemodbus} &  /demo/LINUXTCP/tcpmodbus & Open-Source traces\cite{open_source_for_modbus_dns} & 7 \\
    S7comm & binary & control & snap7\cite{snap7}  & /examples/cpp/x86\_64-linux/server & Open-Source traces\cite{open_source_for_s7} & 5 \\
    Ethernet/IP & binary & control & OpENer\cite{EIP}  & /bin/posix/src/ports/POSIX/OpENer & /bin/posix/src/ports/POSIX/OpENer  & 4 \\
    DNP3.0 & binary & control &  automatak-dnp3\cite{dnp3}  & /cpp/examples/outstation/outstation-demo & Open-Source traces\cite{open_source_for_dnp3}\cite{Igbe2017deterministic} & 3 \\
    FTP & text & network & LightFTP\cite{LightFTP}  & /Source/Release/fftp & Open-Source traces\cite{open_source_for_ftp} & 11 \\
    TFTP & mixed & network & tftpd-hpa\cite{tftpd-hpa}  & /usr/sbin/in.tftpd & Open-Source traces\cite{open_source_for_tftp} & 2 \\
    DNS &  mixed & network & dnsmasq\cite{dnsmasq}  & /src/dnsmasq & Open-Source traces\cite{open_source_for_modbus_dns}  & 2 \\
    HTTP & text & network & miniweb\cite{miniweb}  & /miniweb & Command line tool \texttt{curl}\cite{command_line_tool_for_http} & 7 \\
    \hline
    \end{tabular}
    \label{tab:information of datasets}
\end{table*}

\begin{table*}
  \centering
  \caption{Average F1-scores of semantic inference task across 8 protocols achieved by the evaluated PRE tools.}
  \vspace{-1.0pc}
  \setlength{\tabcolsep}{1.6mm}
  \footnotesize
  \begin{tabular}{c|ccc|cccc|cc|cc|}
   \toprule
   \multirow{2}{*}{Protocol}  & \multicolumn{3}{c}{Command} & \multicolumn{4}{c}{Length} & \multicolumn{2}{c}{Checksum} & \multicolumn{2}{c}{Delim}  \\
   \cmidrule(lr){2-4} \cmidrule(lr){5-8} \cmidrule(lr){9-10} \cmidrule(lr){11-12} 
   & \ourmethod  & \Polyglot   & \DynPRE   & \ourmethod  & \Polyglot   & \Tupni   & \BinaryInferno  & \ourmethod   & \Tupni   & \ourmethod  & \Polyglot   \\
   \midrule
    Modbus  & 1.00  & 0.64  & 0.17  & 0.74  & 1.00  & 1.00  &  0.00 & -  & -  & -  & -    \\ 
    S7comm  & 1.00  & 0.00  & 0.00  & 0.44  & 0.25  & 0.25  & 0.00  & -  & -  & -  & -    \\ 
    Ethernet/IP  & 1.00  & 0.00  &  0.50 &  1.00 &  0.00 & 0.00  & 1.00  & -  & -  &  - &  -  \\ 
    DNP3.0  & 0.00  & 0.00  & 0.00  & 1.00  & 1.00  & 1.00  &  1.00 & 1.00  & 1.00  & -  &  -   \\ 
    DNS  & 0.00  & 0.00  & 0.00  & -  & -  &  - & -  & -  & -  & 0.26  & 0.00    \\ 
    FTP  & 0.00  & 0.00  & 0.00  & -  & -  &  - &  - & -  &  - & 0.74  & 0.00    \\ 
    TFTP  & 1.00  & 0.50  & 0.00  & -  &  - &  - & -  & -  & -  & -  &  -  \\ 
    HTTP & 0.00  & 0.01  & 0.00  &  - & -  & -  & -  & -  & -  & 0.78  & 0.67   \\ 
    \hline
   \textbf{\textit{Average}} & 0.50  & 0.14  & 0.08  & 0.80  & 0.56 & 0.56  & 0.50  &  1.0 & 1.0  & 0.59  & 0.22   \\ 
   
   \bottomrule
  \end{tabular}
      \begin{tablenotes}
        \item * ``-'' means that the protocol does not contain the fields with the corresponding semantic function.
        \end{tablenotes}
  \label{tab:RQ2details}
 \end{table*}

\end{document}